\documentclass[useAMS,usenatbib]{mn2e}

\usepackage{amsmath,amssymb,amsfonts,latexsym}
\usepackage[dvips]{graphicx}
\usepackage{longtable}
\usepackage{fixltx2e}
\usepackage{lscape}
\usepackage{verbatim}
\usepackage{multicol}
\usepackage[english]{babel}

\title{On the pulse--width statistics in radio pulsars.\\ I. Importance of the interpulse emission}

\author[Krzysztof Maciesiak, Janusz Gil and Val\'{e}rio A. R. M. Ribeiro]{Krzysztof Maciesiak$^{1}${\thanks{E-mail:jezyk@astro.ia.uz.zgora.pl}}, Janusz Gil$^{1}$ and Val\'{e}rio A. R. M. Ribeiro$^{2}$\\
$^{1}$Kepler Institute of Astronomy, University of Zielona G\'{o}ra, Lubuska 2, 65-265 Zielona G\'{o}ra, Poland\\
$^{2}$Astrophysics Research Institute, Liverpool John Moores University, Twelve Quays House, Egerton Wharf, Birkenhead, CH41 1LD}
\begin{document}

\date{Accepted . Received ; in original form }

\pagerange{\pageref{firstpage} -- \pageref{lastpage}} \pubyear{2010}

\maketitle

\label{firstpage}

\begin{abstract}
We performed Monte Carlo simulations of different properties of pulsar radio emission, such as: pulsar periods, pulse--widths, inclination angles and rates of occurrence of interpulse emission (IP). We used recently available large data sets of the pulsar periods $P$, the pulse profile widths $W$ and the magnetic inclination angle $\alpha$. We also compiled the largest ever database of pulsars with interpulse emission, divided into the double--pole (DP--IP) and the single--pole (SP--IP) cases. We identified 31 (about 2\%) and 13 (about 1\%) of the former and the latter, respectively, in the population of 1520 normal pulsars. Their distribution on the $P-\dot{P}$ diagram strongly suggests a secular alignment of the magnetic axis from the originally random orientation. We derived possible parent distribution functions of important pulsar parameters by means of the Kolmogorov--Smirnov significance test using the available data sets ($P$, $W$, $\alpha$ and IP), different models of pulsar radio beam $\rho=\rho(P)$ as well as different trial distribution functions of pulsar period $P$ and the inclination angles $\alpha$. The best suited parent period distribution function is the log--normal distribution, although the gamma function distribution cannot be excluded. The strongest constraint on derived model distribution functions was the requirement that the numbers of interpulses generated by means of Monte Carlo simulations (both DP--IP and SP--IP cases) were exactly (within $1\sigma$ errors) at the observed level of occurrences. We found that a suitable model distribution function for the inclination angle is the complicated trigonometric function which has two local maxima, one near 0$^{\circ}$ and the other near 90$^{\circ}$. The former and the latter implies the right rates of IP, occurrence, single--pole (almost aligned rotator) and double--pole (almost orthogonal rotator), respectively. It is very unlikely that the pulsar beam deviates significantly from the circular cross-section. We found that the upper limit for the average beaming factor $f_b$ describing a fraction of the full sphere (called also beaming fraction) covered by a pulsar beam is about 10\%. This implies that the number of the neutron stars in the Galaxy might be underestimated.
\end{abstract}


\section{Introduction}
Although the number of pulsars discovered recently in modern search campaigns increased enormously, the observed pulsar population is still a small fraction of the neutron star population in the Galaxy. Most of them will never be detected as radio pulsars due to misalignment of their beams with our line--of--sight (l--o--s). However, many of those whose beams point towards the Earth still await detection in future, more sensitive pulsar surveys. Therefore, a more or less complete knowledge about Galactic pulsar population can be obtained only by means of statistical considerations. Statistical studies of the pulse--width in mean profiles of radio pulsars is an important tool for investigations of the geometry of pulsar radiation. One especially important parameter that can be derived from such studies is the inclination angle between the magnetic and the spin pulsar axes. Early studies were carried out by \citet{hp69}, \citet{rs72, rs73}, \citet{b76} and Manchester \& Lyne (1977, hereafter \citet{ml77}). Since the amount of the available data was small, these papers suffered from problems of small number statistics. A more complete work was performed by \citet{p79} and \citet{lm88}, who analyzed samples of about 200 pulse--width data measured near 400 MHz. Although the database used in these papers was quite rich, the pulse--width measurements were contaminated by the interstellar scattering dominating at low radio frequencies. More recently Gil \& Han (1996; hereafter \citet{gh96}) compiled a new database of 242 pulse--widths $W_{10}$ (corresponding to about 10 per cent of the maximum intensity) measured at a higher radio frequency (near 1.4 GHz), which was relatively unbiased compared to the lower frequency data. GH96 used their pulse--width database to perform Monte Carlo simulations in an attempt to derive the distribution statistics of pulsar periods, pulse--widths, magnetic inclination angles and rates of interpulses. By comparing the simulated and observed (or observationally derived) quantities they concluded that the observed distribution of the inclination angles resembles a sine function following from the flat (random) distribution in the parent population, and that the probability (beaming fraction) of observing a pulsar was about 0.16. GH96 also pointed out that the rates of interpulse occurrence should be considered as an important aspect of pulsar population studies.

On the other hand, Tauris \& Manchester (1998, hereafter \citet{tm98}) using a different method based on an analysis of the indirectly derived polarization position angles and magnetic inclination angles concluded that the observed distribution of the latter is cosine--like rather than sine--like as suggested by GH96. They also obtained the beaming fraction $0.10\pm 0.02$, considerably lower than 0.16 obtained by GH96. TM98 pointed out a likely source of this discrepancy, namely the incorrect assumption used by GH96 that the observed distribution and the parent distribution of pulsar periods are similar. Recently, Zhang, Jiang \& Mei (2003, hereafter \citet{zjm03}) followed the Monte Carlo simulation scheme developed by GH96. ZJM03 argued that both the parent distribution function and the observed distribution of pulsar periods can be modelled by the gamma function but with different values of the free parameters, and their Monte Carlo simulations included searching for a 2-D grid of these parameters. As a result, ZJM03 concluded that indeed the cosine--like distribution (suggested by TM98) is much more suitable to model the inclination angles in the parent pulsar population than the flat distribution (suggested by GH96). They argued that the most plausible parent distribution is a modified cosine function (see Section 2.1.1), which has a peak around $25^\circ$ and another weaker peak near $90^\circ$. They also obtained the beaming fraction $\sim 0.12$, consistent with the result of TM98.

As emphasized by ZJM03 in the conclusions of their paper, neither they nor TM98 considered potentially important constraints related to the observable interpulse emission. Kolonko et al. (2004, KGM04 hereafter) corrected this shortcoming of analysis of ZJM03 and included the rates of occurrences of the interpulse emission, divided into categories of single--pole and double--pole origin. In this paper we follow the scheme developed by KGM04 but with a few important improvements. We used much richer databases of pulsar periods, pulse--widths and interpulse occurrences. Moreover, we use broader spectrum of trial distribution functions in our Monte Carlo simulations.

\section{Geometry of the pulsar radiation}\label{sec.geometry}
\begin{figure}
\begin{center}
\vspace{30pt}
\includegraphics[width=8cm,height=5.5cm,angle=0]{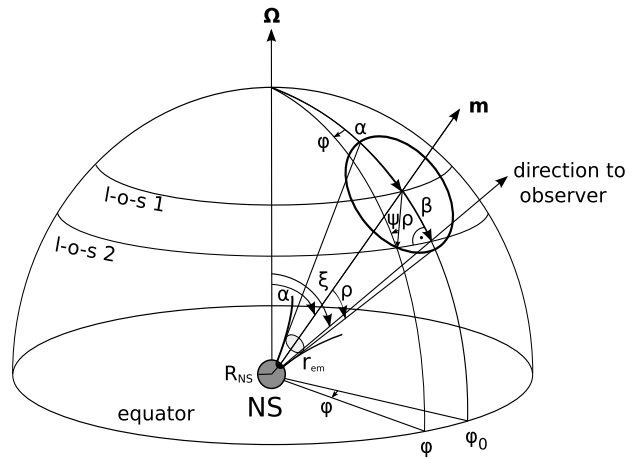}
\caption{Geometry of the pulsar radiation presented schematically on the ''celestial'' hemisphere centred on the neutron star (NS) with the radius $R_{NS}$. The polar cap is marked as a black area on the shadow NS surface. The radio emission region is marked at the altitude $r_{em}$. The fiducial plane $\varphi_0=0$ contains the rotation $\mathbf{\Omega}$ and the magnetic $\mathbf{m}$ axes as well as the observer's direction. The following angles are marked: the longitudinal phase $\varphi$ measured from the fiducial phase $\varphi=0$, the inclination angle $\alpha$ between the magnetic $\mathbf{m}$ and the spin $\mathbf{\Omega}$ axes, the impact angle $\beta$ of the closest approach of the observer to the magnetic axis, the observer's angle $\xi=\alpha+\beta$, the opening angle of the radiation beam $\rho$ and the polarisation position angle $\psi$. Two exemplary line--of--sights are marked: l--o--s 1 passing through the magnetic axis $(\beta=0)$ and l--o--s 2 corresponding to a grazing impact angle $\beta\sim\rho$.
\label{Fig.1}}
\end{center}
\end{figure}

The basic condition for the pulsar to be detected is that it should be bright enough for sensitivity of the observing system used in the radio observatory. In this paper we assume that this condition is always satisfied (some consequences of this assumption are discussed in Appendix \ref{sec.appendix.luminosity} (in the on--line materials)) and our main concern is geometrical detection conditions. Pulsar can be detected if its narrow beam sweeps through the observer. We use this geometrical detection method in our Monte Carlo simulations, without taking into account the intrinsic pulsar luminosity (although we briefly discuss this problem in Section \ref{sec.discussion}). However, we restrict our parameter space to quantities that should not be strongly affected by the luminosity problem.

The geometry of pulsar radiation is schematically shown in Fig. \ref{Fig.1}. The observed pulse--width $W$ (ignoring dispersive and scattering broadening; see Section \ref{sec.pulse.width}) depends on the intrinsic angular radius of the beam $\rho$, the inclination angle $\alpha$ and the impact angle $\beta$. Purely geometrical pulse--width $W_l$ on the $l$-th level of the maximum intensity of the profile is
\begin{eqnarray}
W_l=2\cdot\varphi_l= &\nonumber\\ &4\cdot\arcsin\left\{\left[\frac{\sin\left(\left(\rho_l+\beta\right)/2\right)\cdot\sin\left(\left(\rho_l-\beta\right) /2\right)}{\sin\alpha\cdot\sin\left(\alpha+\beta\right)}\right]^{1/2}\right\},
\label{eq.w.gil81}
\end{eqnarray}
\citep{gil81}. For the impact angle $\beta=0^{\circ}$ (corresponding to a very rare situation when the line--of--sight cuts through the centre of the beam) and the inclination angle $\alpha=90^{\circ}$ (corresponding to the case of the orthogonal rotator), we get $W=2\rho$. For non--orthogonal rotator this gives a very well known approximation $W=2\rho/cos\;\alpha$. Thus, roughly speaking the pulse--width is proportional to the angular radius of the beam $\rho$. We would like to emphasize that Eq. (\ref{eq.w.gil81}) assumes symmetry of the pulsar beam (and thus symmetry of the pulse--width but not necessarily the pulse shape) with respect to the fiducial phase $\varphi_0$ (Fig. \ref{Fig.1}). We paid special attention to selecting pulsars with relatively low dispersion measure $DM$, free from broadening features (see Section \ref{sec.pulse.width}) that would introduce significant asymmetry into pulsar profiles.

\subsection{Probability density distribution functions}\label{sec.prob.density.funct}
The main aim of this work is to carry out statistical studies of emission properties of the normal radio pulsars. On the one hand, we have samples of different measurable (directly or indirectly) parameters obtained for a large number of pulsars. On the other, we can simulate these parameters and compare the simulated values with the measured ones. In each case the important question we try to answer is what the statistical distributions of these parameters are in the observed sample, and more generally in the whole pulsar population. Below we consider possible trial distribution functions of selected pulsar parameters. We present the so-called parent trial distribution functions, which should be distinguished from the observed distributions.

\subsubsection{Inclination angle $\alpha$ and impact angle $\beta$}
In an arbitrary coordinate system the spin axis of the neutron star and the viewer's line--of--sight are both randomly chosen uniformly on the surface of the sphere, resulting in an isotropic distribution of both. The correlated probability distribution of the observer's angle $\xi=\alpha+\beta$ relative to the spin axis is given by
\begin{equation}
f(\xi)=sin(\xi)=sin(\alpha+\beta),
\label{eq.xi}
\end{equation}
where $\alpha$ and $\beta$ are the inclination and the impact angles\footnote{Range of the observer's angle $\xi$, inclination angle $\alpha$ and impact angle $\beta$ is as follows: $0\le\xi\le\pi$, $0\le\alpha\le\pi/2$ and $-\pi/2\le\beta\le\pi/2$. The observer's $\xi$ angle is uniformly distributed on a sphere, however the inclination angle $\alpha$ is drawn from one of the trial probability density functions (Eqs. (\ref{eq.flat}) -- (\ref{eq.cosZJM})). Thus, the impact angle $\beta$ distribution is not uniform but depends on distribution of the inclination angle. Let us keep in mind that for the observed pulsars the impact angle $\beta \le \rho$, where $\rho$ is the angular width of the emission beam (Fig. \ref{Fig.1}). Although distributions of both $\alpha$ and $\beta$ are unknown, their sum is uniformly distributed on a sphere.}, respectively (Fig. \ref{Fig.1}). However, the actual distribution of the inclination angle $\alpha$ may depend on number of unknown factors. Therefore, we considered a number of trial probability density functions for the parent distribution of $\alpha$, including flat distribution
\begin{equation}
f(\alpha)=\frac{2}{\pi},   \label{eq.flat}
\end{equation}
sine function
\begin{equation}
f(\alpha)=\sin\;\alpha,   \label{eq.sin}
\end{equation}
or cosine function
\begin{equation}
f(\alpha)=\cos\;\alpha.   \label{eq.cos}
\end{equation}
Apart from the simple functions presented above some more complicated probability density functions were considered as well. As a parent distribution function of the inclination angle $\alpha$ GH96 proposed the following function
\begin{equation}
f(\alpha)=\frac{5}{4}+\cos\left(\frac{5}{2}\;\alpha\right).   \label{eq.cosGH}
\end{equation}
Later ZJM03 proposed even more complicated function
\begin{equation}
f(\alpha)=\frac{0.6}{\cosh(3.5(\alpha-0.43))}+\frac{0.15}{\cosh(4.0(\alpha-1.6))}   \label{eq.cosZJM}
\end{equation}
which they called the modified cosine function. This function is characterised by two local maxima, around $\alpha\approx25^{\circ}$ and another weaker one around $\alpha\approx90^{\circ}$. Interestingly ZJM03 did not consider rates of interpulse occurrence in their paper and their complicated cosine function was introduced to improve modelling of the parent inclination angles by means of Monte Carlo simulations. Later, KGM04 demonstrated that this function is also responsible for the occurrence of the proper amount of interpulses. We do not confirm this conclusion in the present paper, using much larger databases of the observed pulsar parameters and interpulses. We find that the most suitable parent distribution is that proposed by GH96 (\ref{eq.cosGH}).

\subsubsection{Period $P$}
\citet{gh96} showed that the distribution function of the observed 516 pulsars with periods $0.05 < P < 4.2$ s can be fitted\footnote{Fit of this function to the observed data was obtained with the least square method, minimalising $\displaystyle\sum_{i}^{} \left[N(x_i)-\mathcal{N}_0 \int_{x_i}^{x_{i+1}} x^{a-1}e^{-x}dx \right]^2$. They found values of free parameters $m$ and $a$ as well as the normalising constant $\mathcal{N}_0$ (details in GH96).} by the gamma function
\begin{equation}
f(P)=\mathcal{N}_0 \:x^{a-1}e^{-x},  \label{eq.gamma2}
\end{equation}
where $\mathcal{N}_0$ is the normalisation constant, $x=P/m$ and, $m$ and $a$ are values of the free parameters. \citet{zjm03} used this function to fit much larger sample of the 1165 period values from the same period range and obtained good result. They argued that the parent distribution of periods is different from the observed distribution and it is very convenient to use the gamma function with the values of $m$ and $a$ treated as free parameters of the model. We will use this approach, generalized by including a number of other trial probability density functions, like the Lorentz function
\begin{equation}
f(P)=\frac{\mathcal{C}_0}{1+\left(P-x_0\right)^2/a^2_0},   \label{eq.lorentz}
\end{equation}
where $\mathcal{C}_0$ is the normalisation constant, $x_0$ and $a_0$ are free parameters. Another trial function that we considered was the Gauss function representing normal distribution
\begin{equation}
f(P)=\frac{1}{\sigma\sqrt{2\pi}}\exp\left( \frac{-(P-x_0)^2}{2\sigma} \right),   \label{eq.gauss}
\end{equation}
where $x_0$ and $\sigma$ are free parameters. We also tried the log--normal distribution
\begin{equation}
f(P)=\frac{1}{P\sigma\sqrt{2\pi}}\exp \left[ \frac{-\left(\log P - x_0\right)^2}{2\sigma ^2} \right]   \label{eq.lognor}
\end{equation}
(suggested by \citet{lorimer06}), where $x_0$ and $\sigma$ are free parameters of period probability distribution function whose logarithm is normally distributed.

\subsubsection{Opening angle $\rho$}
The opening angle (radius) of the pulsar beam can be calculated from the pulse--width $W$ if $\alpha$ and $\beta$ angles are known, either from the polarisation data \citep{mt77} or from the width of the core component (using Eq. (\ref{eq.w.core}) taken from Rankin (1990; hereafter \citet{r90})), or both. Inverting Eq. (\ref{eq.w.gil81}) one obtains the opening angle $\rho$ as a function of $\alpha$, $\beta$ and $W$ \citep{gil84}
\begin{equation}
    \rho_l = 2 \sin^{-1}\left[\sin\alpha\;\sin(\alpha+\beta)\;\sin^2\frac{W_l}{4}+\sin^2\left(\frac{\beta}{2}\right) \right]^{1/2}.
\label{eq.rho.from.eq.gil}
\end{equation}

\citet{lm88} were the first who applied this equation to a large number of 10 per cent pulse--width data $W_{10}$ measured at 408 MHz. They argued that $\rho_{10}$(408 MHz) $\approx 6^{\circ}.5 P^{-1/3}$, which scaled to the 1.4 GHz was
\begin{equation}
\rho_{10}=5^{\circ}.8\;P^{-1/3}.   \label{eq.rho.lm}
\end{equation}
Biggs (1990) reanalysed the same sample of data and argued that
\begin{equation}
\rho_{10}=5^{\circ}.6\;P^{-1/2}.   \label{eq.rho.biggs}
\end{equation}
\citet{r93a} analysed a large sample of pulse--widths data taken at different frequencies in different world radio observatories over a period of several years. She has interpolated all available data to frequency of about 1 GHz and divided them into different profile classes (according to \citet{r83}). In each class she obtained a bimodal $\propto P^{-1/2}$ opening angle distribution. This result clearly indicated that pulsar beams consist one or two coaxial cones centred on the magnetic axis, with the opening angle $\rho$ of each cone following $P^{-1/2}$ period dependence. Gil, Kijak \& Seiradakis (1993; GKS93) and \citet{k94} have confirmed this result at frequency 1.4 GHz, using data obtained with the Effelsberg 100 m radiotelescope. Instead of dividing pulsars into different classes to reveal the bimodal $P^{-1/2}$ distribution of $\rho$, \citet{gks93} performed a careful error analysis and rejected all data subject to large errors (broadening the apparent distribution). As a result they obtained that for given period $P$, the opening angle $\rho$ can have two possible values:
\begin{equation}
\rho_{10} = \left\{
\begin{array}{l l}
6^{\circ}.3\;P^{-1/2}\\
4^{\circ}.9\;P^{-1/2}
\end{array} \right.
\label{eq.rho.gks}
\end{equation}
(see Fig. 2 in \citet{gks93}). \citet{k94} obtained exactly the same result, using an independent method for both the pulse--width measurements and error analysis. Examining Fig. 2 in \citet{gks93} we can notice that the inner cone with $\rho=4^{\circ}.9\;P^{-1/2}$ seems to be preferred at shorter periods $P<0.7$ s, while the outer cone with $\rho=6^{\circ}.3\;P^{-1/2}$ dominates at longer periods $P>1.2$ s. However, the exact model of transition between cones is not known. This observational feature is crucial, and it has to be taken into account in the statistical analysis to calculate the pulse--width in the synthetic population. We use Eq. (\ref{eq.rho.gks}) in two model variants:\\
a) based on Fig. 2 in \citet{gks93} we established the period value $P=0.7$ s below which the inner cone ($4^{\circ}.9\;P^{-1/2}$), and above this value the outer cone ($6^{\circ}.3\;P^{-1/2}$), is always chosen,\\
b) like in case a) but below period $P=0.7$ s there is a 20 per cent chance to choose the outer cone and an 80 per cent chance for the inner cone.\\

It is worth noting that Eq. (\ref{eq.rho.gks}) was derived by GKS93 by means of geometrical analysis of a large number of conal profiles. We believe that it describes well the low intensity pulse-width measurements used in this paper.

\subsection{Interpulse emission}\label{sec.geom.interpulses}
At the time of writing the manuscript of this paper there were 1520 normal pulsars (with periods longer than 20 ms) known. In nearly 3 per cent of them the so-called interpulse (IP) emission could be identified, by which we understand features separated by about 180$^{\circ}$ (possible deviation could amount to about 40 per cent) from the main pulse (MP). The canonical lighthouse pulsar model naturally predicts the occurrence of interpulses. In this model two beams are collimated along the open lines of dipolar magnetic field. When the inclination angle $\alpha\approx90^{\circ}$ (almost orthogonal rotator) the observer can detect both beams, associated with two opposite magnetic poles. This is the so-called double--pole interpulse model (hereafter DP--IP). In this case both pulse components are clearly separated (by about 180$^{\circ}$ of longitude) and there is not any kind of low level emission between them. Duty cycles of each component are small, typically several per cent of the pulsar period. Another possibility of generating the interpulse is described by the so-called single--pole model (SP--IP hereafter). This model requires a small inclination angle $\alpha$ (almost aligned rotator). In the SP--IP case pulse--widths are much broader than in DP--IP model, to the extent that they often fill the entire or most of the pulsar period (360$^{\circ}$). Even if both components are separated, usually there is a low intensity bridge of emission between them. The first version of this model (\citet{rl68}) assumes that MP and IP occur when the observer's line--of--sight cuts the wide hollow cone of radiation twice (ML77) at a distance of about 180$^{\circ}$ of longitude (Fig. \ref{Fig.C1} (Appendix \ref{sec.appendix.figures} in the on--line materials)). In the other version of SP--IP model (Fig. \ref{Fig.C2} (Appendix \ref{sec.appendix.figures} in the on--line materials)) the line--of--sight stays in a pulsar beam for the entire pulsar period and MP and IP correspond to cuts through two nested conical beams or through the arrangement of the core beam surrounded by the cone (\citet{gil83}; hereafter G83, \citet{gil85}). In the latter version the separation between MP and IP is naturally close to 180$^{\circ}$, and it does not depend on the observational frequency, while in the former version these properties are not natural.

In addition to information about profile shape and/or widths one can usually use polarisation angle (PA) curves to distinguish between different kinds of interpulses. In case of DP-IP (almost orthogonal rotators) swings of PA are steep across both MP and IP components, while for SP-IP cases PA curve is typically flat over the entire profile (including MP, IP and weak emission bridge between them).\\

\subsection{Beam shape}
It is commonly assumed that the pulsar beam is circular in shape. Some authors however consider elliptical shapes with meridional compression or equatorial elongation. For example, \citet{nv83} argued that the pulsar beam is elongated with the ratio of north--south (N--S) to east--west (E--W) dimension depending on pulsar period as
\begin{equation}
    R\approx1.8 P^{-0.65},
\label{eq.nv83}
\end{equation}
so the effect is predominant at shorter periods $P<0.25$ s. On the other hand, \citet{biggs90} and \citet{kinnon93} considered the opposite tendency of the beam compression in meridional direction, with the ratio of minor (N--S) to major (E--W) ellipse axes depending on the inclination angle $\alpha$
\begin{equation}
    R\approx\cos\frac{\alpha}{3}\sqrt{\cos\left(\frac{2}{3}\alpha\right)},
\end{equation}
so this effect is predominant at large $\alpha$ close to $\pi/2$.
Equations (\ref{eq.w.gil81}) and (\ref{eq.rho.from.eq.gil}) are independent from the beam shape (if symmetry to the fiducial plane is conserved). The opening angle is always described by Eq. (\ref{eq.rho.from.eq.gil}) or by its observational representations (Eq. (\ref{eq.rho.lm}), (\ref{eq.rho.biggs}) or (\ref{eq.rho.gks})). Taking into account ellipse geometry the following equation could be shown
\begin{equation}
      r=\sqrt{\rho_0^2+\beta^2\left(\frac{1}{R^2}-1\right)}
\label{eq.r.compr}
\end{equation}
and
\begin{equation}
    rR=\sqrt{R^2\rho_0^2+\beta^2\left(1-R^2\right)}.
\label{eq.rR.compr}
\end{equation}
Equation (\ref{eq.rR.compr}) describes circular beam when $R=1$ and $r=\rho$. In this paper we argue that pulsar beam shape is circular or almost circular (a possible ellipticity can not be excluded but if it exists it is very small).

\subsection{Detection conditions}\label{sec.detect.cond}
For simplicity let us consider one hemisphere with $0\le\alpha\le\pi/2$ (Fig. \ref{Fig.1}). Pulsar is detectable geometrically when the observer's l--o--s passes through its beam (independent of the actual beam shape). For circular or almost circular beam the detection condition is
\begin{equation}
    \vert\beta\vert < \rho_0,
\label{eq.det.c}
\end{equation}
where $\rho_0$ is the opening angle of the beam corresponding to the last open field lines. The above condition is valid for the main pulse emission. For interpulse emission within DP--IP model (almost orthogonal rotator $\alpha \sim \pi /2$) the detection condition is
\begin{equation}
    \rho_0 > \pi-2\alpha-\beta.
\label{eq.det.c.dp}
\end{equation}
In the SP--IP model, in which both MP and IP originate from single magnetic pole (almost aligned rotator $\alpha\sim 0$), we will consider two different versions of this model. In the first MP and IP represent two cuts through one conical beam (ML77) and the detection condition is
\begin{equation}
    \rho_0 > \sqrt{2\alpha^2+\beta^2+2\alpha\beta}.
\label{eq.det.c.spm}
\end{equation}
This version is presented in more details in Fig. \ref{Fig.C1}. The second version corresponds to the case when the l--o--s remains inside the beam for the entire pulsar period, so the occurrence of both MP and IP result from the internal beam structure in the form of nested hollow--cones (G83). The detection condition for this case is
\begin{equation}
\rho_0 > 2\alpha+\beta.   \label{eq.det.c.spg}
\end{equation}
This version is presented in more details in Fig. \ref{Fig.C2}.

For elliptical pulsar beams the detection conditions expressed by equations (\ref{eq.det.c}), (\ref{eq.det.c.dp}), (\ref{eq.det.c.spm}) and (\ref{eq.det.c.spg}) transform into equations (\ref{eq.det.e}), (\ref{eq.det.e.dp}), (\ref{eq.det.e.spm}) and (\ref{eq.det.e.spg}), respectively, presented below:
\begin{equation}
\vert\beta\vert < [R^2\rho_0^2+\beta^2(1-R^2)]^{1/2},    \label{eq.det.e}
\end{equation}

\begin{equation}
    rR > \pi-2\alpha-\beta,
\label{eq.det.e.dp}
\end{equation}

\begin{equation}
    rR > \sqrt{2\alpha^2+\beta^2+2\alpha\beta}
\label{eq.det.e.spm}
\end{equation}

\begin{equation}
    rR > 2\alpha+\beta.
\label{eq.det.e.spg}
\end{equation}
In the case of SP--IP model, beside detection conditions (Eq. (\ref{eq.det.c.spm}) or Eq. (\ref{eq.det.e.spm})) it is also important to apply some kind of morphological definition to avoid a danger of classifying just a broad double--peaked profiles as interpulses. By carefully reviewing our interpulse database we decided that the interpulse case has to have at least 100$^{\circ}$ of longitude separation between the component peaks, that is
\begin{equation}
    100^{\circ} < \Delta \varphi_{sep}^{IP-MP} < 260^{\circ}.
\label{eq.sp.definition}
\end{equation}

\subsection{Beam edge correction}\label{sec.beam.edge}
\begin{figure}
%\vspace{10pt}
\begin{center}
\includegraphics[width=8cm,height=8cm,angle=0]{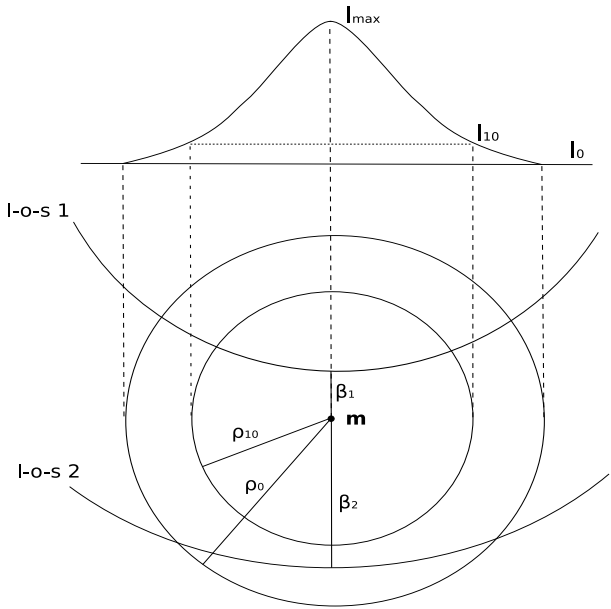}
\caption[]{Schematic representation of the gaussian envelope of the beam intensity distribution. Two concentric circles centred on the magnetic axis $\mathbf{m}$ with radii $\rho_0$ and $\rho_{10}$ represent two intensity levels: edge of the beam and 10\% of the maximum intensity in the beam centre, respectively. Two lines of sight trajectories l--o--s 1 and l--o--s 2 corresponding to different impact angles $\beta_1$ and $\beta_2$ are marked. One can notice that in the case of l--o--s 1 the detection independent of the actual values of $\rho$ ($\rho_0$ or $\rho_{10}$) in the detection condition (Eq. (\ref{eq.det.c})) is possible because $\beta<\rho_{10}<\rho_0$. However, in the case of l--o--s 2 the condition $\beta<\rho_{10}$ does now guarantee the detection, and new detection condition $\beta<\rho_{0}=1.1\rho_{10}$ should be introduced.
\label{Fig.2}}
\end{center}
\end{figure}

KGM04 used the detection condition $\rho>\vert\beta\vert$ (Eq. (\ref{eq.det.c})), where $\rho$ was approximated by $\rho_{10}$ obtained from observations (e.g. from models Eqs. (\ref{eq.rho.lm}) -- (\ref{eq.rho.gks})). Such an approach was justified only by the fact that it is difficult to measure pulse--width at the level lower than about 10\% of the maximum intensity. It seems that this condition is too restrictive from the geometrical point of view, because the intensity level corresponding to the $\rho_{10}$ beam radius could be much higher than the one that should be adopted as the edge of the beam. A simple model of the beam envelope is presented in Fig. \ref{Fig.2}. The two circles with different radii $\rho_0$ and $\rho_{10}$ correspond to two different intensity levels with respect to the maximum intensity level $I_{max}$ in the centre of the beam, namely 10 \% and low intensity level corresponding to the beam edge. It is important to notice that the maximum of the observed profile (upper panel) does not generally correspond to the maximum intensity of the beam and in reality depends on the impact angle $\beta$. Thus $I_{10}$ does not necessarily correspond to the cut of the beam at 10\% of its maximum (it is just 10\% of the maximum of the observed profile). If we use $\rho=\rho_{10}$ in detection conditions we can lose $\sim10\%$ pulsars with intensity less than 0.1 of the absolute maximum of the beam. To improve this situation, let us consider Tables 1a and 1b in \citet{gk93}. These tables are very useful since they present the pulse--width measurements at 10\% and 1\% maximum intensity of the profile for a group of pulsars. The latter one can be considered as being close to the level representing the edge of the beam. The values of the inclination angle $\alpha$ and the impact angle $\beta$ are known, so using Eq. (\ref{eq.rho.from.eq.gil}) we can calculate corresponding values of $\rho_{10}$ and $\rho_1$, which for available data give the mean value close to $1.1$. If we now adopt $\rho_0=\rho_1$ we can estimate the average value of the ratio of $\rho_0/\rho_{10}=1.1$, which we can use as the edge beam correction in detection conditions (Eqs. (\ref{eq.rho.lm}) -- (\ref{eq.rho.gks}) and Eqs. (\ref{eq.r.compr}) -- (\ref{eq.det.e.spg})).

\section[]{Databases}

\subsection{Periods}\label{sec.database.periods}
In this paper we consider only the so-called normal radio pulsars. All magnetars, millisecond pulsars and pulsars in binary systems are excluded, since they constitute different evolutionary groups on the $P-\dot{P}$ diagram. Most of pulsar periods $P$ are available in the ATNF\footnote{http://www.atnf.cisiro.au/research/pulsar/psrcat} catalogue \citep{manchester05}. Our database was consecutively updated according to new papers publishing new pulsar discoveries (e.g. \citet{janssen09}, \citet{keith09}, \citet{keith10}).\\

\begin{figure}
\vspace{10pt}
\begin{center}
\includegraphics[width=7cm,height=14cm,angle=0]{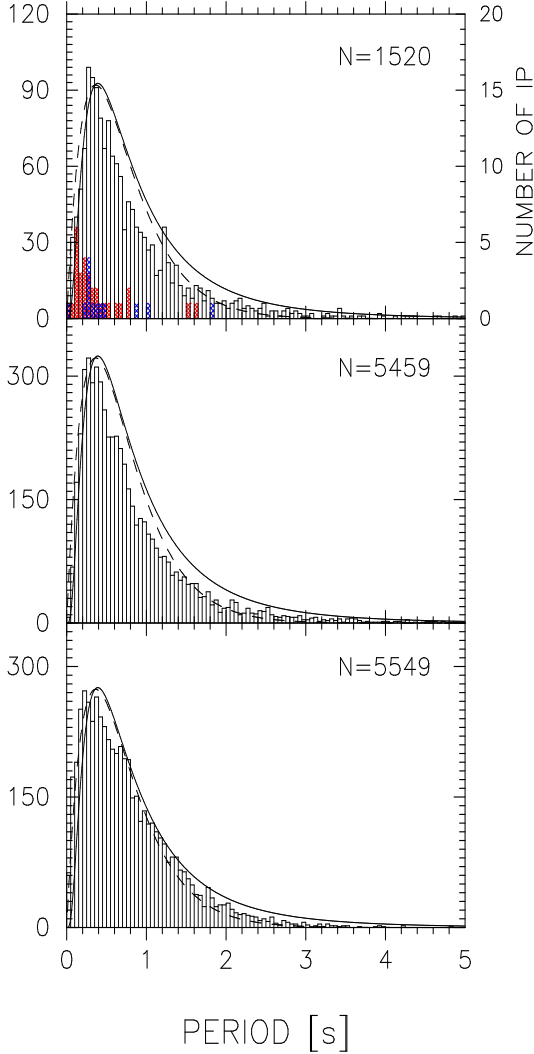}
\caption{Formal fits of the log--normal (Eq. (\ref{eq.lognor})) and the gamma (Eq. (\ref{eq.gamma2})), solid and dashed line respectively, distribution functions to the histogram of 1520 observed (upper panel) and simulated (lower panels) pulsar periods. The ordinate represents the number of observed (upper panel) or simulated (lower panels) pulsars. The upper panel shows also the distribution of interpulses according to the right-hand side scale, where blue and red colour corresponds to SP-IP and DP-IP cases, respectively, (as in Fig. \ref{Fig.6})). Distribution of the simulated observed pulsar periods generated with the log--normal (middle panel) and the gamma (lower panel) parent period distribution functions with parameters $x_0=-0.30$, $\sigma=0.80$ and $m=0.43$, $a=2.12$, respectively. In both simulation cases the inclination angle distribution function was that of GH96 (Eq. (\ref{eq.cosGH})). The parameter $N$ describes the number of observed and simulated periods.
\label{Fig.3}}
\end{center}
\end{figure}

At present there are 1830 known radio pulsars, with periods that range from 1.4 ms to 11.8 s. Following arguments expressed by \citet{kgm04} we consider only those with periods ranging from 20 ms to 8.5 s, both in our period database and in Monte Carlo simulations. The upper limit of 8.5 s is set by PSR J2144-3933 (\citet{southern1} and \citet{young99}) with the longest period observed in radio wavelengths, whereas the lower limit we set at 20 ms. The estimates of initial pulsar period range from 14 ms \citep{migliazzo02} to 140 ms \citep{kramer03a}. If the actual shortest period is indeed 14 ms instead of being close to 20 ms, then we miss only two normal pulsars, which is statistically irrelevant. All in all, our period database contains 1520 normal pulsars, 355 more than in the KGM04 database. The distribution of all selected periods is presented in the upper panel of Fig. \ref{Fig.3}. This sample can be best fitted by the log--normal function expressed by Eq. (\ref{eq.lognor}) with parameters $x_0=-0.30$ and $\sigma=0.80$, although gamma function (Eq. (\ref{eq.gamma2})) with parameters $m=0.43$ and $a=2.12$ also gives a reasonable fit (see Table  \ref{tab.2} for comparison).

\subsection{Pulse--widths}\label{sec.pulse.width}
\begin{figure}
\vspace{10pt}
\begin{center}
\includegraphics[width=7cm,height=14cm,angle=0]{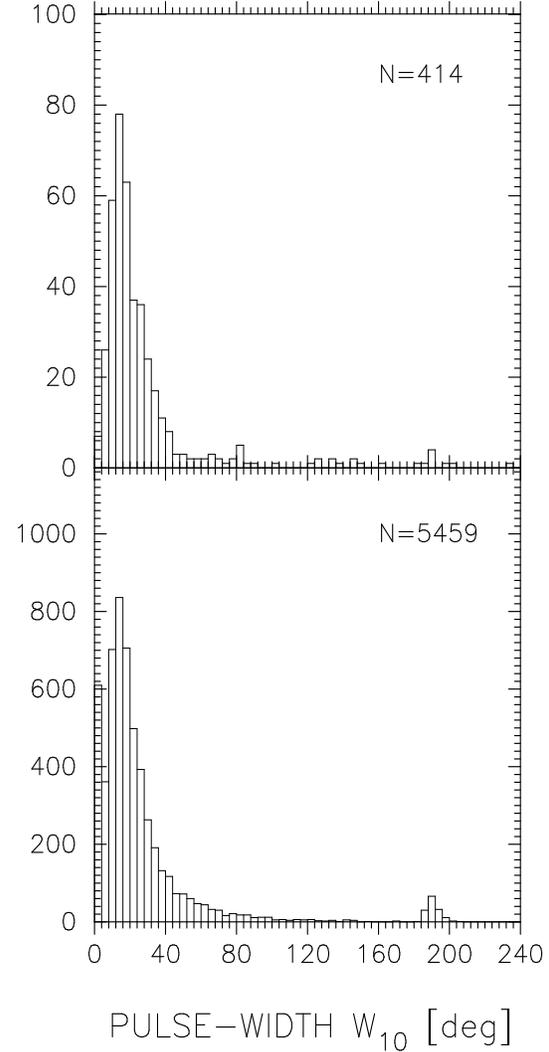}
\caption{Distribution of 414 pulse--widths $W_{10}$ measurements for pulsars with $DM\le150$ pc cm$^{-3}$ (upper panel). The distribution of simulated values is presented in the lower panel. The simulations were performed using the log--normal period distribution (Eq. (\ref{eq.lognor})) and the GH96 inclination angle distribution (Eq. (\ref{eq.cosGH})), as this combination gives the best results of the K-S significance test. The parameter $N$ describes the number of observed and simulated periods.
\label{Fig.4}}
\end{center}
\end{figure}
In our simulations we attempt to reproduce the observed distribution of pulse--widths measured at 10\% of the maximum intensity level. The observed pulse--width $W$ depends on a number of geometrical (discussed in Section \ref{sec.geometry}) and non-geometrical factors (like scattering, dispersion smearing, sampling or receiver time constant). The influence of the emission geometry (inclination angle $\alpha$, impact angle $\beta$, opening angle $\rho(P)$) on the observed pulse--width $W$ (Eq. (\ref{eq.w.gil81})) is dominant, however some influence of non-geometrical factors cannot be neglected.

Both \citet{gh96} and \citet{kgm04} used database that contained about 240 (242 and 238, respectively) pulse--widths that were carefully selected to avoid broadening effects. The trailing part of many profiles with large dispersion measure $DM$ suffer from broadening caused by scattering of radio waves on free electrons in ISM. Our simulations do not take into account scattering, so our sample should be limited to pulsars whose profiles are not significantly broadened. The easiest selection method is to limit the values of dispersion measure but the question is the maximum permissible $DM$ that warrants it? To answer this question we analysed the Fig. 7 in \citet{cl03}. We found that the limiting value of $DM$ is about 150 pc cm$^{-3}$ and we present our reasoning below.

The typical duty cycle of normal pulsars is $\sim5\%$, which for our shortest period (20 ms) gives a pulse window of about 1 ms. We assumed arbitrarily the value of 10 per cent (0.1 ms) of this value as the maximum possible broadening due to high $DM$. According to Fig. 7 in \citet{cl03} this corresponds to dispersion measure value $DM\sim150$ pc cm$^{-3}$. Of course, the number of pulsars with period close to 20 ms in the whole sample is comparatively small, so probably the $DM$ limit can be slightly higher. Increasing the maximum $DM$ value to 170 pc cm$^{-3}$ would extend the number of pulse--widths by less than 10 per cent, which seems to be irrelevant in our statistical analysis. However, for $DM\sim200$ pc cm$^{-3}$ the pulse broadening would be about 1 ms, which is evidently too much. Thus, we construct our new pulse--width database using the value of 150 pc cm$^{-3}$ as the limit for $DM$ in our sample.

The new database contains pulse--widths used by GH96 and KGM04, extended by pulse--widths from Swinburne Intermediate -- Latitude Pulsar Survey \citep{swinburne}, Parkes Southern Pulsar Survey I -- III (\citet{southern1}, \citet{southern2}, \citet{southern3}), Parkes Multibeam Pulsar Survey I -- VI (\citet{manchester01}, \citet{morris02}, \citet{kramer03b}, \citet{hobbs04}, \citet{faulkner04}, \citet{lorimer06} and \citet{keith09}).
Parkes Southern Pulsar Survey I and II campaigns were carried out at 436 MHz and others on 1.4 GHz, so we had to scale their pulse--widths to 1.4 GHz\footnote{Scaling factor is 0.89 (= 1.4 GHz/0.436 GHz)$^{-0.1}$. This factor was obtained in the following way: the pulse--width is roughly proportional to the angular radius of the beam $\rho$, which is proportional to square root of the emission altitude $r_{em}^{1/2}$, which in turn depends on observational frequency $\nu$ ($r_{em}=(400\pm80)\:\text{km}\; \nu_{\text{GHz}}^{-0.26\pm0.09}\;\dot{P}_{-15}^{\:0.07\pm0.03}\;P^{\:0.30\pm0.05}$; \citet{kijak97,kijak98}). Thus, the pulse--width can be roughly scaled with the observational frequency $\nu$ as $\nu^{-0.1}$.}. After rejecting of all magnetars, milliseconds and binary pulsars 768 normal pulsars remained, among which 414 had $DM\le150$ pc cm$^{-3}$. Thus, our new pulse--width database contains 414 measurements, including 190 from GH96/KGM04 database. Their distribution is presented in the upper panel of Fig. \ref{Fig.4}.

\subsection{Inclination angles}
\begin{figure}
\vspace{35pt}
\begin{center}
\includegraphics[width=7cm,height=14cm,angle=0]{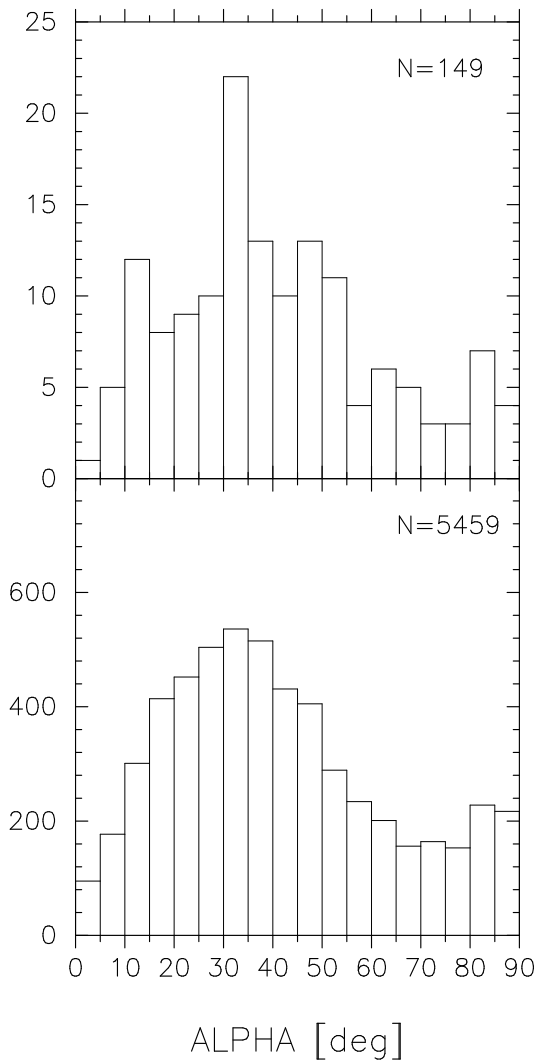}
\caption{The upper panel shows the distribution of 149 inclination angles $\alpha$ derived from the pulsar polarisation characteristics (data taken from \citet{r93a} and \citet{r93b}. The histogram of simulated inclination angles is presented in the lower panel. The simulations were performed using the log--normal period distribution (Eq. (\ref{eq.lognor})) and the GH96 inclination angle distribution (Eq. (\ref{eq.cosGH})), as this combination gives the best results of the K-S significance test.
\label{Fig.5}}
\end{center}
\end{figure}

The inclination angle $\alpha$ between the magnetic and the spin pulsar axes is a parameter which cannot be observed but can be derived indirectly. One method is based on the RVM and polarisation data (\citet{lm88}, \citet{r90}, \citet{r93a,r93b}). It is well know that this method is effective only in a small number of cases with very broad highly polarized profiles. Another, much more effective method was proposed by \citet{r90}, in which $\alpha$ was derived from the width of the core component measured at the level of 50\% maximum intensity ($W_{50}^{core}$)
\begin{equation}
    \alpha=sin^{-1}\left(2^{\circ}.45 P^{-1/2} / W_{50}^{core} \right).
\label{eq.w.core}
\end{equation}
In our simulation we used database of 149 inclination angles compiled by \citet{r93a,r93b}. Their distribution is presented in Fig. \ref{Fig.5} (upper panel). Unfortunately, the inclination angle database is the only one which was not developed since Rankin's compilation. We would like to advertise here that the validity of the above Equation (\ref{eq.w.core}) will be examined statistically in the forthcoming Paper II.

\subsection{Interpulses}
As a result of discovery of large number of pulsars in recent multibeam surveys a number of cases with interpulse emission also increased from 14 in \citet{taylor93} catalogue to 27 in Weltevrede and Johnston (2008a; thereafter WJ08a) up to 44 presented in this paper. The new interpulse database was derived from the sample of 1520 normal pulsars. This database (Table \ref{tab.1}) contains 31 double--pole interpulsars (DP--IP) and 13 single--pole interpulsars (SP--IP), which is 2.039\% and 0.855\% of the total number of normal pulsars, respectively\footnote{More precisely, the rates of occurrence of SP--IP and DP--IP are $(0.86 \pm 0.24)\%$ and $(2.04 \pm 0.37)\%$, respectively. The errors were estimated $1\sigma=\sqrt{N}/N_{tot}$, which for SP--IP is $\sqrt{13}/1520=0.24\%$ and for DP--IP is $\sqrt{31}/1520=0.37\%$ ($N_{tot}=1520$ is the number of normal pulsars from which the interpulses were extracted).}. A clear division into DP and SP interpulses was possible in most cases, based on the pulse profile morphology and/or polarization angle variations (see Section \ref{sec.geom.interpulses}). It is important to note that the ratio of $N_{SP-IP}/N_{IP}=0.30$ and $N_{DP-IP}/N_{IP}=0.70$ is almost the same as in \citet{kgm04} (0.36 and 0.64). The total percentage $\sim2.90\%$ of interpulsars in the sample of normal pulsars is also unchanged. Thus, it seems that our new database is quite representative for interpulse occurrences in normal pulsars.

To be consistent with our selection of pulsar periods we excluded all millisecond and other recycled pulsars from both the total pulsar sample and from the sample of pulsars with interpulses (this was not done in \citet{gh96}, who incorrectly estimated rates of interpulse occurrences as a result of this mistake).

\begin{figure*}
  \vspace*{5pt}
\includegraphics[width=16cm,height=15cm,angle=0]{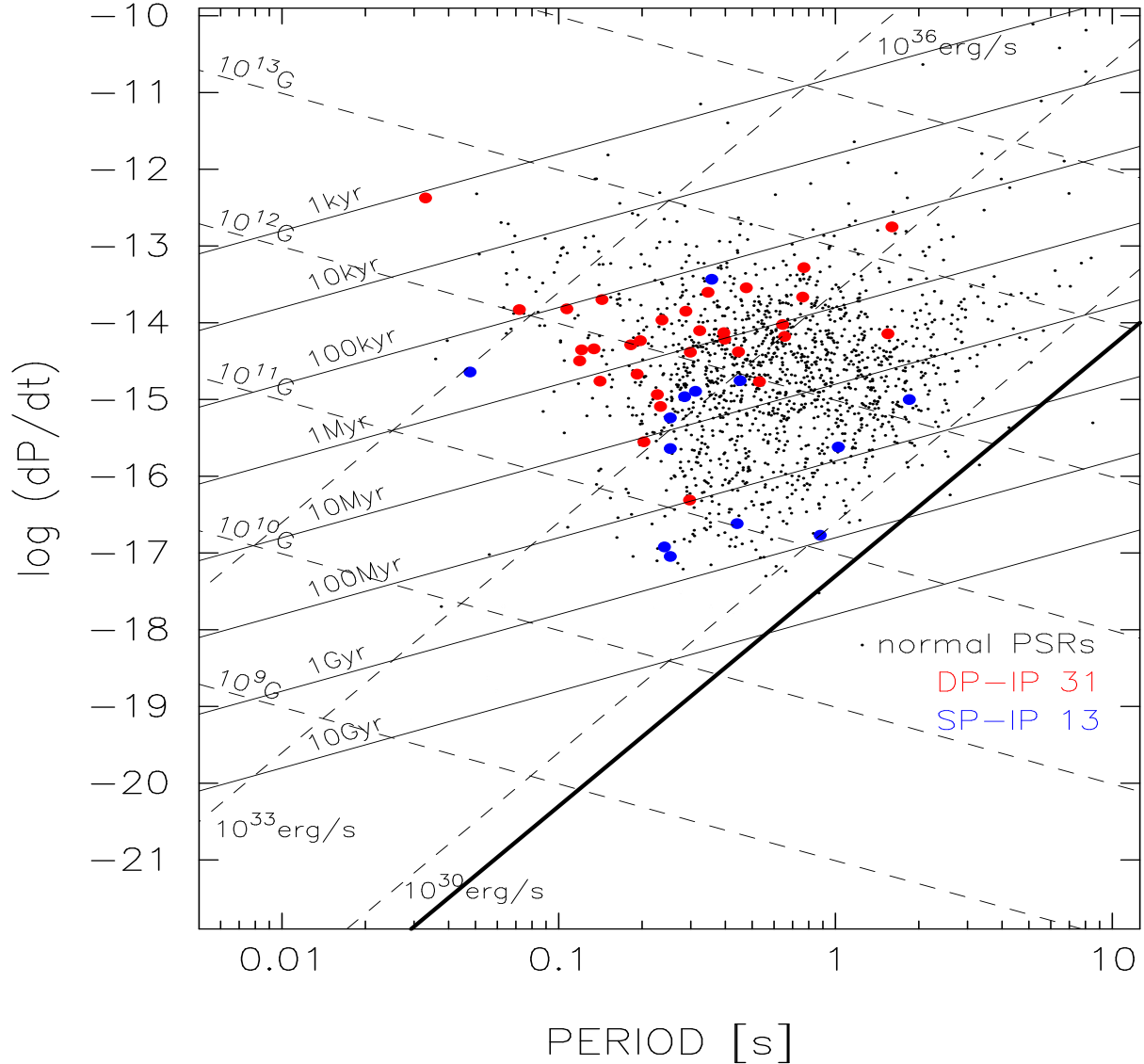}
\vbox to 20mm{\vfil\caption{Diagram $P-\dot{P}$ for normal (see Section \ref{sec.database.periods}) pulsars, including 44 cases with interpulse emission (see Table \ref{tab.1}). Black dots represent 1476 normal pulsars, while red and blue circles correspond to 31 double--pole (DP--IP) and 13 single--pole (SP--IP) interpulsars, respectively. Lines of constant magnetic field $B$, characteristic age $\tau$, spin--down luminosity $\dot{E}$ are shown. The death line ($10^{29}$ erg s$^{-1}$) derived by \citet{conto06} is marked by the thick line. It is easy to notice that SP--IP cases are generally located much closer to the death line than DP--IP cases. \label{Fig.6}} \vfil}
\end{figure*}

All interpulsars (see Table \ref{tab.1}) are marked on the $P-\dot{P}$ diagram presented in Fig. \ref{Fig.6}.

\begin{table*}
 \centering
 \begin{minipage}{145mm}
  \caption{Table of 44 known interpulses divided into double--pole (DP--IP) and single--pole (SP--IP) cases. Bibliographic marks are following: T/K -- \citet{kgm04} using \citet{taylor93}, WJ -- \citet{wj08b}, A -- \citet{southern3}, M02 -- \citet{morris02}, K03 -- \citet{kramer03b}, H -- \citet{hobbs04}, L -- \citet{lorimer06}, R -- \citet{ribeiro08}$^{\dag}$, M01 -- \citet{manchester01}, J -- \citet{janssen09}, K09 -- \citet{keith09}, K10 -- \citet{keith10}, C -- \citet{camilo09}, N -- new interpulse identified in this work using data from H and L. \label{tab.1}}
  \begin{tabular}{|r|l|l|r|c|c|c|c|c|}
\hline
\multicolumn{1}{|c|}{} & \multicolumn{1}{|c|}{}  & \multicolumn{1}{|c|}{} & \multicolumn{1}{|c|}{}& \multicolumn{1}{|c|}{}  & \multicolumn{1}{|c|}{Ratio of} & \multicolumn{1}{|c|}{Separation}  & \multicolumn{1}{|c|}{SP-IP} & \multicolumn{1}{|c|}{} \\

\multicolumn{1}{|c|}{No.} & \multicolumn{1}{|c|}{Name J}  & \multicolumn{1}{|c|}{Name B} & \multicolumn{1}{|c|}{Period}& \multicolumn{1}{|c|}{$\dot{P}$} &
\multicolumn{1}{|c|}{amplitudes}& \multicolumn{1}{|c|}{MP and IP}& \multicolumn{1}{|c|}{or}    & \multicolumn{1}{|c|}{Bibliography} \\

\multicolumn{1}{|c|}{} & \multicolumn{1}{|c|}{}  & \multicolumn{1}{|c|}{} & \multicolumn{1}{|c|}{[s]}& \multicolumn{1}{|c|}{$[10^{-15}]$} &
\multicolumn{1}{|c|}{IP/MP} & \multicolumn{1}{|c|}{$[^{\circ}]$}  & \multicolumn{1}{|c|}{DP-IP} & \multicolumn{1}{|c|}{} \\\hline

1. & J$0534+2200$ & B$0531+21$& 0.033 & $423$ & 0.6   & 145 & DP & T/K \\
2. & J$0826+2637$ & B$0823+26$& 0.531 & $1.7 $& 0.005 & 180 & DP & T/K \\
3. & J$0828-3417$ & B$0826-34$& 1.849 & $0.99$& 0.1   & 180 & SP & T/K \\
4. & J$0908-4913$ & B$0906-49$& 0.107 & $15.2$& 0.24  & 176 & DP & T/K \\
5. & J$0953+0755$ & B$0950+08$& 0.253 & $0.2$ & 0.012 & 210 & SP & T/K \\
6. & J$1057-5226$ & B$1055-52$& 0.197 & $5.8 $& 0.5   & 205 & DP & T/K \\
7. & J$1302-6350$ & B$1259-63$& 0.048 & $2.3$ & 0.75  & 145 & SP & T/K \\
8. & J$1705-1906$ & B$1702-19$& 0.299 & $4.1$ & 0.15  & 180 & DP & T/K \\
9. & J$1722-3712$ & B$1719-37$& 0.236 & $10.9$& 0.15  & 180 & DP & T/K \\
10.& J$1739-2903$ & B$1736-29$& 0.323 & $7.9 $& 0.4   & 180 & DP & T/K \\
11.& J$1825-0935$ & B$1822-09$& 0.769 & $52.3$& 0.05  & 185 & DP & T/K \\
12.& J$1851+0418$ & B$1848+04$& 0.285 & $1.1 $& 0.2   & 200 & SP & T/K \\
13.& J$1932+1059$ & B$1929+10$& 0.227 & $1.2$ & 0.018 & 170 & DP & T/K \\
14.& J$1946+1805$ & B$1944+17$& 0.441 & $0.02$& 0.005 & 175 & SP & T/K \\\hline\hline
15.& J$0905-5127$ &           & 0.346 & $24.9$& 0.059 & 175 & DP & WJ \\
16.& J$1126-6054$ & B$1124-60$& 0.203 & $0.03$& $\sim$0.1&174& DP& WJ \\
17.& J$1611-5209$ & B$1607-52$& 0.182 & $5.2  $& $<$0.1& 177 & DP & WJ \\
18.& J$1637-4553$ & B$1634-45$& 0.119 & $3.2 $& $\sim$0.1&173&DP & WJ \\\hline\hline
19.& J$1549-4848$ &          & 0.288   &$14.1$&$\sim$0.3&180& DP & A\\\hline\hline
20.& J$1806-1920$&   &0.880& $0.017$ & $\sim$1.0 & $\sim$136 & SP & M02 \\
21.& J$1828-1101$& &0.072  & $14.8 $ & $\sim$0.3 & 180 & DP & M02 \\
22.& J$1913+0832$& &0.134  & $4.6  $ & $\sim$0.6 & 180 & DP & M02 \\\hline\hline
23.& J$0834-4159$& &0.121  & $4.4   $& $\sim$0.25 & 171 & DP & K03\\
24.& J$1713-3844$& &1.600  & $177.4 $& $\sim$0.25 & 181 & DP & K03\\\hline\hline
25.& J$1843-0702$& &0.192  & $2.1   $& $\sim$0.44 & 180 & DP & H\\
26.& J$1852-0118$& &0.452  & $1.8   $& $\sim$0.4 & 144 & SP & H\\\hline\hline
27.& J$1808-1726$& &0.241  & $0.012 $& $\sim$0.5 & $\sim$223 & SP & L\\
28.& J$1849+0409$& &0.761  & $21.6  $& $\sim$0.5 & 181 & DP & L\\ \hline\hline
29.& J$0831-4406$& & 0.312 & $1.3   $& $\sim0.05$ & 244 & SP & R -- K03 \\
30.& J$1107-5907$& & 0.253 & $0.09  $& $\sim0.2$ & 191 & SP & R -- L \\
31.& J$1424-6438$& & 1.024 & $0.24  $& $\sim0.12$ & 223 & SP & R -- K03 \\
32.& J$1613-5234$& & 0.655 & $6.6   $& $\sim$0.28 & 175 & DP & R -- M01 \\
33.& J$1627-4706$& & 0.141 & $1.7   $& $\sim0.13$ & 171 & DP & R -- L \\
34.& J$1637-4450$& & 0.253 & $0.58  $& $\sim0.26$ & 256 & SP & R -- L \\
35.& J$1842+0358$& & 0.233 & $0.81  $& $\sim0.23$ & 175 & DP & R -- L\\
36.& J$1915+1410$& & 0.297 & $0.05  $& $\sim$0.21& 186 & DP & R -- L\\ \hline\hline
37.& J$0842-4851$& B$0840-48$&0.644  & $9.5$ & $\sim$0.14 & 180 & DP & N -- H\\
38.& J$1413-6307$& B$1409-62$&0.395  &7.434&$\sim0.04$ & $\sim170$ & DP & N -- H\\
39.& J$1737-3555$& B$1734-35$&0.398  &6.12&$\sim0.04$ & $\sim180$ & DP & N -- H\\
40.& J$1903+0925$&   &0.357  & $36.9$& $\sim0.19$ & $\sim240$ & SP & N -- L\\\hline\hline
41.& J$2047+5029$&   &0.446  & $4.2$ & $\sim$0.6  & 175 & DP & J\\ \hline\hline
42.& J$1244-6531$&   &1.547  & $7.2$ & $\sim$0.3  & 145 & DP & K09\\
43.& J$0627+0706$&   &0.476  & $29.9^{\dag\dag}$ &$\sim$0.2  & 180 & DP & K10 \\\hline\hline
44.& J$2032+4127$&   &0.143  & $20.1$ &$\sim$0.18 & 195 & DP & C \\\hline
\multicolumn{9}{|l|}{$^{\dag}$ Items 29 -- 36 correspond to interpulses identified by \citet{ribeiro08} in the data of K03, L and M01.} \\
\multicolumn{9}{|l|}{$^{\dag\dag}$ Unpublished value provided by Michael Keith (private communication).}\\

\end{tabular}
\end{minipage}
\end{table*}

As one can see from this figure DP--IP cases (represented by red dots) lie above 10 Myr line. This means that almost orthogonal rotators are rather young pulsars. On the other hand, SP--IP cases (blue dots) lie between $\sim$5 Myr and $\sim$500 Myr lines. Thus, the almost aligned rotators have a tendency to be older pulsars. In fact, the average values of $P$, $\dot{P}$, $\tau$ and $B$ are 0.51 s, $3.49\times10^{-15}$, $1.55\times10^8$ yr and $6.88\times10^{11}$ G for SP--IP cases and for DP--IP cases are 0.39 s, $2.86\times10^{-14}$, $4.64\times10^6$ yr and $2.17\times10^{12}$ G. This strongly implies a secular alignment of the magnetic axis towards the spin axis, in agreement with the two humped distribution function of parent inclination angles expressed by Eq. (\ref{eq.cosGH}). To some extent this is an observational support for results of simulation of the inclination angle $\alpha$ evolution made by WJ08a. These authors argued that the magnetic axis is likely to align from a random distribution at birth with a timescale of $\sim10^7$ years. It is not clear whether this conclusion is fully consistent with Eq. (\ref{eq.cosGH}), although the alignment of the magnetic axis seems to be reproduced well in Fig. \ref{Fig.6}.

\section{Monte Carlo simulations of conal emission}\label{sec.mc}
We performed Monte Carlo simulation of the pulse--widths $W_{10}$, pulsar periods $P$ and inclination angles $\alpha$ following a technique developed by KGM04, with a number of important improvements:
\begin{enumerate}
\item [1.] We used new, generally more numerous databases of pulsar periods (1520 as compared with 1165 in KGM04), pulse--widths (414 versus 238) and interpulses (44 versus 14). The inclination angle database was unchanged with respect to KGM04.
\item [2.] The pulse--widths measurements were selected with additional criterion that $DM\le 150$ pc cm$^{-3}$. This allows avoiding a significant pulse broadening contaminating pulse--width measurements at the low intensity level.
\item [3.] The log--normal period distribution proposed by \citet{lorimer06} was added to a set of possible trial functions.
\item [4.] Beam edge correction in detection conditions was introduced in the form $\rho_0=1.1 \rho_{10}$ where $\rho_{10}$ is expressed by Eqs. (\ref{eq.rho.lm}) -- (\ref{eq.rho.gks}). For more details see Section \ref{sec.beam.edge}.
\item [5.] The separation between MP and IP within the ML77 version of SP--IP model is computed exactly from the detection geometry. As one can see from Tab. \ref{tab.1}, the minimum (maximum) separation between MP and IP is about 100$^{\circ}$ (260$^{\circ}$). On the other hand, we have the empirical relation for component peaks
\begin{equation}
\rho_s = \left\{
\begin{array}{l l}
3^{\circ}.7\;P^{-1/2}\;\; \textrm{for} P>0.7 \textstyle{s} \\
4^{\circ}.6\;P^{-1/2}\;\; \textrm{for} P\leq0.7 \textstyle{s}
\end{array} \right.
\label{eq.rho.sep}
\end{equation}
found by \citet{gks93} (see their Fig. 3), which can be used in Monte Carlo simulations. While using the wide hollow cone version of the SP--IP model (ML77) one can have a problem judging whether the pulse represents just a broad profile or it is already an interpulse case. Thus, beside satisfying the detection condition for the SP--IP (Eq. (\ref{eq.det.c.spm}) or Eq. (\ref{eq.det.e.spm})) the simulated pulse--width $W(\rho_s)$ should be larger than 100$^{\circ}$ (smaller than 260$^{\circ}$).
\end{enumerate}

The first test for our software was to reproduce results from KGM04 (using their databases). Results of this test are presented in Table \ref{tab.kgm.results} (Appendix \ref{sec.appendix.tables} in the on--line materials). After making sure that our programs work correctly, we started new Monte Carlo simulations. They involve a number of subsequent steps described below:
\begin{enumerate}
\item [1.] Generate the inclination angle $\alpha$ as a random number with one of the trial parent probability density function $f(\alpha)$ corresponding to Eqs. (\ref{eq.flat}) -- (\ref{eq.cosZJM}).

\item [2.] Generate the pulsar period of $P$ as a random number with the one of the trial parent probability density function $f(P)$ corresponding to Eqs. (\ref{eq.gamma2}) -- (\ref{eq.lognor}).

\item [3.] Generate the observer angle $\xi$ as random number with the parent probability density function $f(\xi)=sin\;\xi$.

\item [4.] Calculate the impact angle $\beta=\xi-\alpha$.

\item [5.] Calculate the beam opening angle $\rho_{10}(P)$ at the level of 10\% of the maximum intensity with one of the trial distribution function (Eqs. (\ref{eq.rho.lm}) -- (\ref{eq.rho.gks})).

\item [6.] Calculate the opening angle $\rho_0$ corresponding to the edge of the beam $\rho_0=1.1\rho_{10}$.

\item [7.] Check the detection conditions for each simulated object (Eqs. (\ref{eq.det.c}) -- (\ref{eq.det.c.spg}) for circular and Eqs. (\ref{eq.det.e}) -- (\ref{eq.det.e.spg}) for elliptical beam). If the pulsar is detected then it is added the to total number of detected pulsars $N_{det}$.

\item [8.] Calculate the pulse--width $W_{10}$ according to Eq. (\ref{eq.w.gil81}) for each set of parameters $\rho_{10}(P)$, $\alpha$ and $\beta$ of the detected pulsar.

\item [9.] For each pulsar detected in a simulation run check the detection conditions for occurrence of an interpulse. For DP--IP model Eq. (\ref{eq.det.c.dp}) for circular beam or Eq. (\ref{eq.det.e.dp}) for elliptical beam were used. For SP--IP model in ML77 version Eq. (\ref{eq.det.c.spm}) for circular or Eq. (\ref{eq.det.e.spm}) for elliptical beam were used while for G83 version Eq. (\ref{eq.det.c.spg}) for circular or Eq. (\ref{eq.det.e.spg}) for elliptical beam were used. Moreover, for the ML77 version of the SP--IP model, components separation (Eq. (\ref{eq.rho.sep})) is checked whether it is in the range of 100$^{\circ}$ -- 260$^{\circ}$. If an interpulse is detected then it is added to the number of interpulses $N_{DP-IP}$ or $N_{SP-IP}$.

\item [10.] Record all relevant information like inclination angle $\alpha$, period $P$, pulse--width $W_{10}$ and occurrence of the interpulses from one or two magnetic poles.

\item [11.] For each set of the probability density function of the inclination angle $\alpha$, period $P$ and the model of the opening angle $\rho_{10}$, 50000 objects were created (each having a set of parameters such as: $P$, $\alpha$, $\beta$, $\rho$ and $W$). After checking the detection conditions we obtained database of the observed simulated pulsars and distribution of their parameters $\alpha$, $P$ and $W_{10}$ as well as occurrence of the interpulses from one $N_{SP-IP}$ or two $N_{DP-IP}$ magnetic poles. In order to obtain the statistical significance each simulation run was repeated 10 times and the results were averaged.

\item [12.] Judge the statistical significance using the Kolmogorov--Smirnov (K--S) test for the simulated observed pulsars parameters as $\alpha$, $P$ and $W_{10}$ and observed distribution of these parameters. Record the values of $\mathcal{D}(\alpha)$, $\mathcal{P}(\alpha)$, $\mathcal{D}(P)$, $\mathcal{P}(P)$, $\mathcal{D}(W_{10})$ and $\mathcal{P}(W_{10})$, where $\mathcal{D}$ describes the maximum distance between both cumulative distribution functions and $\mathcal{P}$ is the probability that the two compared distributions are drawn from the same parent distribution.

\item [13.] Calculate the beaming fraction defined as $f_b = N_{det}/N_{tot}$ (where $N_{det}$ is the number of detected pulsars in total population of $N_{tot}$ simulated pulsars) and the occurrence of interpulses from one magnetic pole $N_{SP-IP}/N_{det}$ or two magnetic poles $N_{DP-IP}/N_{det}$, respectively.

\item [14.] Change values of free parameters of a given trial period distribution function.

\item [15.] Repeat simulations with all possible combinations of distribution probability functions of $\alpha$, $P$ and $\rho_{10}$.

\item [16.] Search for good solutions satisfying simultaneously all demanded criteria, that is $\mathcal{P}(\alpha)\geq0.5\%$, $\mathcal{P}(P)\geq0.5\%$, $\mathcal{P}(W_{10})\geq0.5\%$ as well as the rates of occurrence of SP--IP and DP--IP are at the observed level $(0.86\pm0.24)\%$ and $(2.04\pm0.37)\%$ (errors correspond to 1$\sigma$ level).
\end{enumerate}

We performed Monte Carlo simulation of conal emission of radio pulsars according to procedure described in items 1--16 listed above. We checked 160 (80 for circular and 80 for elliptical beams) different combinations of trial functions for parent distribution periods $P$, inclination angles $\alpha$ and opening angles $\rho_{10}$. For each combination of the above distribution functions we searched the two-dimensional grid of free parameters of the period distribution function. It gave the total number of $641\:360$ single simulation runs. In all solutions we identified ''detectable'' cases in which all probabilities (of $P$, $\alpha$ and $W_{10}$) were greater than 0.005 (0.5\%) and occurrence of interpulses were at the observed levels ($2.04\pm0.37$)\% for DP--IP and $(0.86\pm0.24)\%$ for SP--IP. Such solutions we called ''a good solution''.

In summary, each simulation run including 50000 detection attempts resulted in about 5500 detections (satisfying geometrical detection conditions (Section \ref{sec.detect.cond})). There were $641\:360$ simulation runs including all possible combinations of the probability distribution functions (Section \ref{sec.prob.density.funct}). Among the resulting $\sim4\times10^9$ geometrical detections we found 827 good solutions (as described in item 16 above). These solutions are listed in Tables B2 -- B7 and 9 best examples with relatively high K--S probability values are shown in Table \ref{tab.2} (items 1a -- 9a). Items 1b -- 9b correspond to tests of the luminosity problem described in the Appendix \ref{sec.appendix.luminosity} in the on--line materials.

\begin{figure}
\vspace{5pt}
\begin{center}
\includegraphics[width=8cm,height=8cm,angle=0]{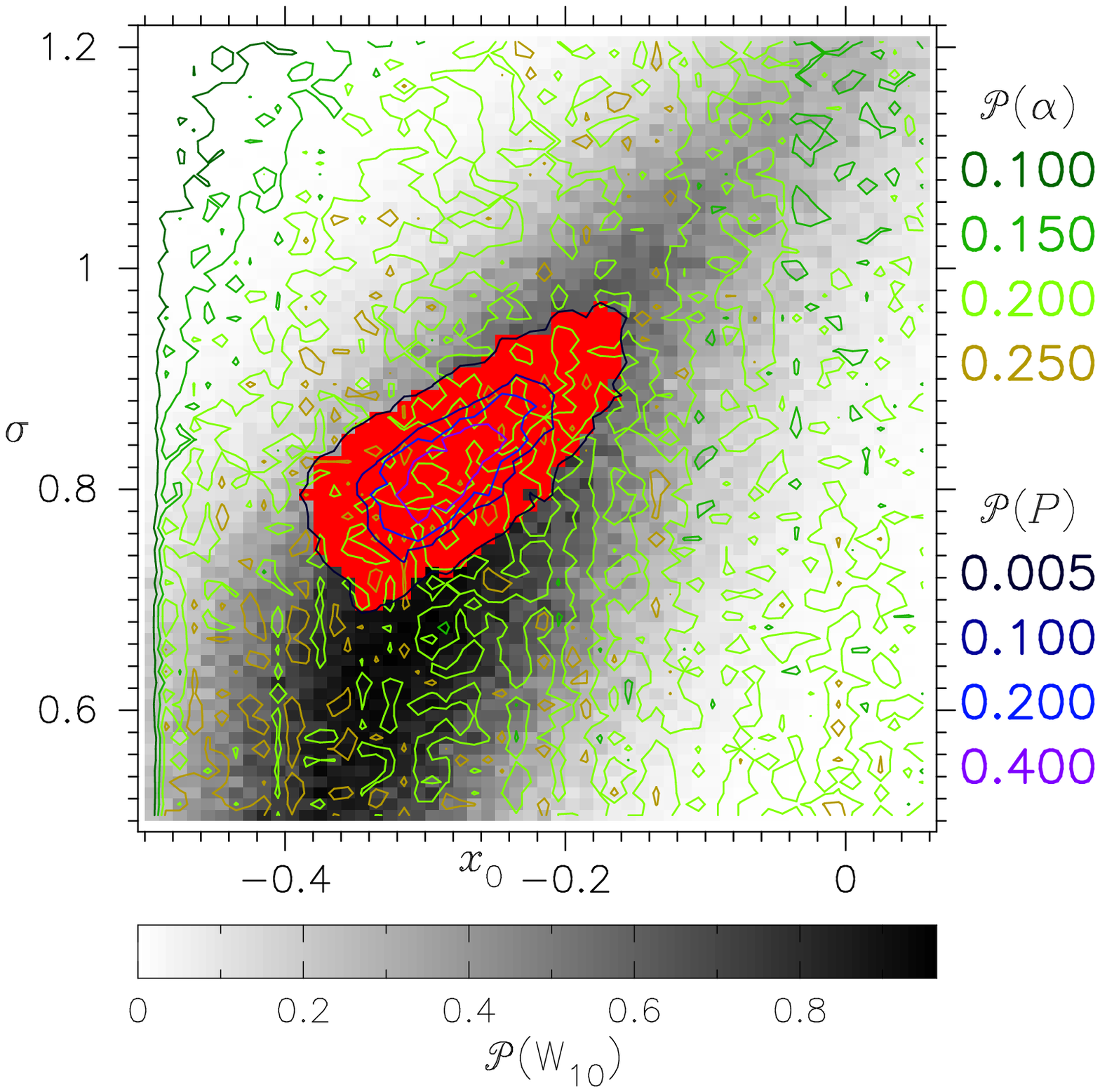}%454
\caption{Colour graphic representation of simulation of the best solution of the Monte Carlo simulations corresponding to the following set of trial probability density function: log--normal (Eq. (\ref{eq.lognor})) for periods, GH96 Eq. (\ref{eq.cosGH}) for inclination angles and Eq. (\ref{eq.rho.lm}) for opening angles. Both axes describe free parameters of the period distribution log--normal function: parameter $x_0$ on the horizontal axis and the parameter $\sigma$ on the vertical axis (these parameters are responsible for the location of the maximum and the width of the log--normal function, respectively). Below the main window the narrow bar with the Kolmogorov--Smirnov probabilities $\mathcal{P}(W)$ for the pulse--width $W_{10}$ coded in shades of grey is presented (the highest value $\sim0.95$ corresponds to K--S probability obtained in this case). On the right hand the scales for K--S probabilities for inclination angles $\mathcal{P}(\alpha)$ and for periods $\mathcal{P}(P)$ are presented by different colours. Red rectangles in the solution area correspond to solutions simultaneously satisfying all constraints (K--S probabilities $\mathcal{P} > 0.005$ for periods, inclination angles and pulse--widths as well as the occurrence of interpulses at the observed levels, i.e. $(0.86\pm0.24)\%$ for SP--IP and $(2.04\pm0.37)\%$ for DP--IP). Thus all acceptable solutions are represented by red rectangles and their distribution describes errors of the log--normal functions $x_0=-0.30^{+0.13}_{-0.09}$ and $\sigma=0.80^{+0.16}_{-0.11}$ (see also Tab.\ref{tab.new.res.454f}).
\label{Fig.7}}
\end{center}
\end{figure}

\begin{figure}
\vspace{5pt}
\begin{center}
\includegraphics[width=8cm,height=8cm,angle=0]{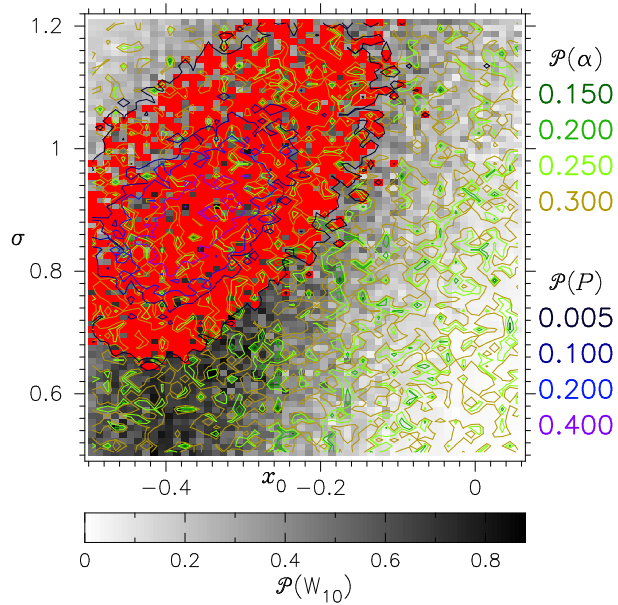}
\caption{Same as in Fig. \ref{Fig.7} but with luminosity problem included in simulations (Appendix \ref{sec.appendix.luminosity}). This figure corresponds to the items 1b and 2b, while Fig. \ref{Fig.7} corresponds to the items 1a and 2a in Table \ref{tab.2}.
\label{Fig.8}}
\end{center}
\end{figure}

\begin{table*}
 \centering
 \begin{minipage}{150mm}
  \caption{Examples of 9 best simulation results for different distribution functions (Eqs. (\ref{eq.cosGH}) -- (\ref{eq.rho.gks})) and circular beams. $\mathcal{D}$ and $\mathcal{P}$ are parameters of the Kolmogorov--Smirnov significance test. For symbol description see item 12 in Section 4. Lines denoted by letters a and b correspond to simulations without and with the luminosity problem included, respectively. Notice that for combination (\ref{eq.cosGH}), (\ref{eq.gamma2}) and (\ref{eq.rho.biggs}) of distribution functions there is only one good solution that could be presented (see Fig. \ref{Fig.C6}).
\label{tab.2}}
  \begin{tabular}{|r|l|c|c|r|r|r|r|r|r|r|r|r|}
\hline\hline
%\tiny
\multicolumn{1}{|r|}{No.} &\multicolumn{1}{|c|}{Distributions} & \multicolumn{2}{|c|}{$f(P)$}& \multicolumn{2}{|c|}{$W_{10}$} &
\multicolumn{2}{|c|}{$P$} & \multicolumn{2}{|c|}{$\alpha$} & \multicolumn{1}{|c|}{\scriptsize{DP-IP}}& \multicolumn{1}{|c|}{\scriptsize{SP-IP}}&\multicolumn{1}{|c|}{$f_b$}\\ \cline{3-10}

\multicolumn{1}{|r|}{} & \multicolumn{1}{|l|}{Equation numbers} &\multicolumn{1}{|c|}{}& \multicolumn{1}{|c|}{} &
\multicolumn{1}{|c|}{$\mathcal{D}$} & \multicolumn{1}{|c|}{$\mathcal{P}$}& \multicolumn{1}{|c|}{$\mathcal{D}$} & \multicolumn{1}{|c|}{$\mathcal{P}$} & \multicolumn{1}{|c|}{$\mathcal{D}$} & \multicolumn{1}{|c|}{$\mathcal{P}$} &
\multicolumn{1}{|c|}{\footnotesize{[\%]}}& \multicolumn{1}{|c|}{\footnotesize{[\%]}}& \multicolumn{1}{|c|}{} \\ \hline \hline

 & & $x_0$ & $\sigma$ & & & & & & & & & \\\cline{3-4}
 1a  & (\ref{eq.cosGH}) (\ref{eq.lognor}) (\ref{eq.rho.lm}) &  -0.30  & 0.80  &  0.032  &  0.821  &  0.018  &  0.813  &  0.089  &  0.196  & 2.10  & 0.89  & 0.108  \\
 1b  &  (\ref{eq.cosGH}) (\ref{eq.lognor}) (\ref{eq.rho.lm})  &  -0.30  & 0.80  &  0.043  &  0.738  &  0.074  &  0.032  &  0.093  &  0.265  & 2.12  & 0.82  & 0.013  \\ \hline
 2a  &  (\ref{eq.cosGH}) (\ref{eq.lognor}) (\ref{eq.rho.lm})  &-0.29  & 0.81  &  0.035  &  0.734  &  0.019  &  0.797  &  0.089  &  0.195  & 2.26  & 0.89  & 0.107  \\
 2b  &  (\ref{eq.cosGH}) (\ref{eq.lognor}) (\ref{eq.rho.lm})  &  -0.29  & 0.81  &  0.042  &  0.745  &  0.068  &  0.066  &  0.080  &  0.431  & 2.01  & 0.76  & 0.013  \\ \hline \hline

 3a  & (\ref{eq.cosGH}) (\ref{eq.lognor}) (\ref{eq.rho.biggs}) &  -0.19  & 0.80  &  0.074  &  0.029  &  0.018  &  0.827  &  0.084  &  0.255  & 2.34  & 1.09  & 0.108 \\
 3b  & (\ref{eq.cosGH}) (\ref{eq.lognor}) (\ref{eq.rho.biggs})   &  -0.19  & 0.80  &  0.066  &  0.341  &  0.062  &  0.122  &  0.095  &  0.294  & 2.01  & 1.06  & 0.013  \\ \hline
 4a  &(\ref{eq.cosGH}) (\ref{eq.lognor}) (\ref{eq.rho.biggs})   & -0.18  & 0.81  &  0.070  &  0.055  &  0.019  &  0.757  &  0.083  &  0.271  & 2.35  & 1.06  & 0.107 \\
 4b  & (\ref{eq.cosGH}) (\ref{eq.lognor}) (\ref{eq.rho.biggs})  &  -0.18  & 0.81  &  0.066  &  0.326  &  0.079  &  0.017  &  0.090  &  0.301  & 2.26  & 0.91  & 0.013  \\ \hline \hline

 5a  & (\ref{eq.cosGH}) (\ref{eq.lognor}) (\ref{eq.rho.gks}b) &  -0.31  & 0.77  &  0.047  &  0.373  &  0.029  &  0.285  &  0.076  &  0.362  & 2.28  & 1.01  & 0.110  \\
 5b  & (\ref{eq.cosGH}) (\ref{eq.lognor}) (\ref{eq.rho.gks}b) & -0.31  & 0.77  &  0.044  &  0.704  &  0.078  &  0.023  &  0.087  &  0.336  & 2.13  & 0.83  & 0.013  \\\hline
 6a  & (\ref{eq.cosGH}) (\ref{eq.lognor}) (\ref{eq.rho.gks}b) &  -0.28  & 0.81  &  0.046  &  0.393  &  0.027  &  0.362  &  0.084  &  0.257  & 2.10  & 1.01  & 0.110  \\
 6b  & (\ref{eq.cosGH}) (\ref{eq.lognor}) (\ref{eq.rho.gks}b) &  -0.28  & 0.81  &  0.043  &  0.723  &  0.089  &  0.006  &  0.087  &  0.335  & 1.98  & 0.81  & 0.014  \\ \hline \hline

& & $m$ & $a$ & & & & & & & & & \\\cline{3-4}
 7a  & (\ref{eq.cosGH}) (\ref{eq.gamma2}) (\ref{eq.rho.lm}) &    0.42  & 2.15  &  0.037  &  0.649  &  0.041  &  0.047  &  0.083  &  0.267  & 2.14  & 1.05  & 0.109  \\
 7b  &  (\ref{eq.cosGH}) (\ref{eq.gamma2}) (\ref{eq.rho.lm}) &   0.42  & 2.15  &  0.051  &  0.546  &  0.094  &  0.006  &  0.083  &  0.402  & 2.12  & 0.74  & 0.013  \\  \hline
 8a  &  (\ref{eq.cosGH}) (\ref{eq.gamma2}) (\ref{eq.rho.lm}) &   0.45  & 2.07  &  0.038  &  0.623  &  0.043  &  0.037  &  0.092  &  0.173  & 2.19  & 0.96  & 0.109  \\
 8b  &  (\ref{eq.cosGH}) (\ref{eq.gamma2}) (\ref{eq.rho.lm}) &   0.45  & 2.07  &  0.049  &  0.576  &  0.104  &  0.005  &  0.088  &  0.343  & 2.17  & 0.76  & 0.013  \\   \hline\hline

 9a  & (\ref{eq.cosGH}) (\ref{eq.gamma2}) (\ref{eq.rho.biggs})  &   0.38  & 2.51  &  0.068  & 0.065  & 0.049   &  0.012  &  0.111  &  0.056  & 2.18  & 1.12  & 0.110  \\
 9b  & (\ref{eq.cosGH}) (\ref{eq.gamma2}) (\ref{eq.rho.biggs})  &  0.38   & 2.51  &  0.055  & 0.470  & 0.108  & 0.001 &  0.105 & 0.169 & 2.11 & 0.885 & 0.014 \\ \hline

\end{tabular}
\end{minipage}
\end{table*}

\section{Discussion}\label{sec.discussion}
Our new statistical analysis is based on a number of new databases that we compiled from the recently published data. We followed methodology developed by KGM04, which relies on comparison of synthetic and real pulsar data. Our period $P$ database contains 1520 items (we excluded all recycled and binary pulsars) in the range of 0.02 -- 8.51 seconds. This database includes 355 more pulsars than recently analysed database compiled by KGM04. We also compiled a new database of pulse--widths $W_{10}$ measured at 10\% intensity level. This database contains 414 items, which is 176 more than that of KGM04. There are many more pulse--width measurements available these days, but our database is restricted to pulsars with $DM<150$ pc cm$^{-3}$, which guarantees avoiding significant external pulse broadening.

Most importantly, we created the largest ever database of interpulse occurrence in pulsar emission. Our database contains 44 pulsars (compared with 14 pulsars in \citet{taylor93}/KGM04 database), including 31 cases of DP--IP and 13 cases of SP--IP. Although our IP database is more numerous than any of the previous ones (e.g. KGM04, WJ08a), it seems that the frequencies of occurrence are similar to those occurring in previous smaller databases. In fact, we have 2.90\% of total number of IP cases, divided into 2.04\% DP--IP and 0.86\% SP--IP cases, respectively, in a population of 1520 pulsars. This can be compared with 2.71\%, 1.94\% and 0.78\%, respectively, found in KGM04 database. One can therefore firmly state that there should be about 3\% of IP cases in the population of normal pulsars, including about 2\% and 1\% of DP--IP and SP--IP cases, respectively.

All pulsars from the period $P$ database for which the value of $\dot{P}$ is known are presented on the $P-\dot{P}$ diagram in Fig. \ref{Fig.6}. Black dots represent 1476 normal pulsars (without IP emission) while red and blue dots correspond to 31 DP--IP and 13 SP--IP cases, respectively. It is easy to notice that SP--IP cases are much older than DP--IP cases (with mean characteristic age 155 Myr versus 4.6 Myr, respectively). Also SP--IP cases represent weaker magnetic fields than DP--IP cases ($6.9\times10^{11}$ versus $2.2\times10^{12}$ G). Assuming reasonable that DP--IP and SP--IP cases represent almost orthogonal ($\alpha$ close to $90^{\circ}$) and almost aligned ($\alpha$ close to $0^{\circ}$) rotators, respectively, one can conclude that the distribution of IP cases on the $P-\dot{P}$ diagram presented in Fig. \ref{Fig.6} reveals a secular alignment of the magnetic axis towards the spin axis with a random initial value of the inclination angle. Similar conclusion was reached by \citet{wj08a} and most recently by \citet{young10}, although using different statistical arguments. It is interesting to note that \citet{young10} reached their conclusion without using interpulse pulsars at all, while these objects were crucial for our analysis. Moreover, \citet{young10} argued that the best suitable solution for their simulation is Model II (see their Fig. 9) whereas our simulations seem to prefer their Model III.

The SP--IP cases are very interesting. In our simulations we used two models of SP--IP emission. In the classical model of ML77 (Fig. \ref{Fig.C1}) the MP and the IP components result from two cuts of a very wide hollow emission cone. In this model the separation between MP and IP should be frequency dependent (following a spread of dipolar field lines), and its value being close to 180$^{\circ}$ should be considered accidental \citep{wws07}. An alternative SP--IP model was proposed by G83 (Fig. \ref{Fig.C2}). This model is based on the assumption of the double conal structure of the pulsar beam. The line--of--sight of the nearly aligned rotator cuts one cone for the MP and the other one for the IP emission (see Fig. 8 in \citet{kloumann10} for schematic diagram for PSR B1944+17). This model naturally predicts two important properties: frequency independence of MP -- IP separation (equal to $180^{\circ}$) and existence of the bridge of emission between MP and IP. Interestingly, in our Monte Carlo simulations we detected more cases corresponding to ML77 model than to G83 model. Therefore, if these simulations are correct, most of the 14 SP--IP cases (about 1\% of a total population) should show tendency to frequency dependent separation between MP and IP. Unfortunately, we are not able to verify this conclusion, as most of the available data correspond to a single frequency. However, this should be performed in some suitable project in the future.

A convenient method of graphical representation of each set of trial distribution functions is ''a colour map of solutions'', example of which is presented in Fig. \ref{Fig.7}. More plots of this kind are presented in Appendix \ref{sec.appendix.figures}. Colour contours represent levels of conformity of the observed and the simulated distributions. The ''green'' set of colours is used for inclination angles and the ''blue'' one is used for periods. Their legends are presented on the right--hand side of Fig. \ref{Fig.7}, which describe the corresponding probability levels $\mathcal{P}(\alpha)$ and $\mathcal{P}(P)$. The grey bar below the plot represents conformity of the observed and the synthetic distributions for the pulse--width $W_{10}$ in the form of K--S probability $\mathcal{P}(W)$. The white colour represents zero probability and the black corresponds to the maximum probability occurring in a given plot. Axes of each plot represent the free parameters of period distribution function (i.e. $m$ and $a$ for the gamma function, $x_0$ and $\sigma$ for the log--normal function). The space for good solutions is restricted to the areas where all three probabilities $\mathcal{P}$ simultaneously exceed 0.5\%. That is why in Fig. \ref{Fig.7} and Fig. \ref{Fig.C3} -- \ref{Fig.C6} all good solutions must lay in darkest blue contour. Moreover, the occurrence of interpulsars in this region must be on the observed level (specified at the end of the previous paragraph), which is marked by red rectangles in Fig. \ref{Fig.7}. In fact, the good solutions are represented by red rectangles because they are marked only if all required probabilities occur simultaneously. The numerical details of all good solutions are presented in Tables \ref{tab.new.res.454f} -- \ref{tab.new.res.351f}.

Careful studies of all contour plots (see examples in Fig. \ref{Fig.7} and Fig. \ref{Fig.C3} -- \ref{Fig.C9}) and corresponding numerical solutions (Tables \ref{tab.new.res.454f} -- \ref{tab.new.res.351f}) lead to the following conclusions:
\begin{enumerate}
\item [1.]Solutions satisfying all criteria ($\mathcal{P} > 0.5\%$ for all variables $P$, $\alpha$ and $W_{10}$ as well as the definition and the occurrence of interpulses at the observed levels) were obtained mainly for the log--normal period distribution function (Eq. (\ref{eq.lognor})) with parameters $x_0=-0.30^{+0.13}_{-0.09}$ and $\sigma=0.80^{+0.16}_{-0.11}$ (see Fig. \ref{Fig.7}). Six best examples corresponding to this function are presented in Table \ref{tab.2} (items 1 -- 6). Also the gamma distribution period function (Eq. (\ref{eq.gamma2})) with parameters $m=0.45^{+0.07}_{-0.08}$ and $a=2.04^{+0.34}_{-0.16}$ (see Fig. \ref{Fig.C5}) seems to be quite good, however the corresponding K--S probabilities are lower than in the log--normal distribution period function case. Four best examples are presented in Table \ref{tab.2} (items 7 -- 9). None of the period trial functions (Eqs. (\ref{eq.gamma2}) -- (\ref{eq.lognor})) is perfectly suited to reproduce the observed distributions of our observables. However, the gamma and the log--normal functions are by far the best, with the latter being slightly better, most likely because it reproduces the tail of long periods (see middle panel in Fig. \ref{Fig.3}) very well.

\item [2.] Both the rotation axis and the observer's direction are randomly distributed in space (see Eq. (\ref{eq.xi})).

\item [3.] The only trial distribution function of the inclination angle that satisfies all constrains is the complicated trigonometric function of GH96 (represented by Eq. (\ref{eq.cosGH})). This function has two local maxima, one near 0$^{\circ}$ (almost aligned rotator) and the other near 90$^{\circ}$ (almost orthogonal rotator). We do not find support for the modified cosine function of ZJM03 (Eq. (\ref{eq.cosZJM})), which was derived without taking into account a problem of frequency of occurrence of the IP emission. It is clear that non of the functions represented by Eqs. (\ref{eq.flat}) -- (\ref{eq.cosZJM}) reproduces the observed distribution of simulated variables $P$, $W$ and $\alpha$ as well as the frequency of IP occurrence. However, the function of GH96 represented by Eq. (\ref{eq.cosGH}) suits the best and is recommended as the model function for the parent distribution of the inclination angles.

\item [4.] As a result of simulations performed with suitable parent distribution functions of periods and inclination angles we obtained good solutions for most of the trial model functions for the opening angle. It is not possible to discriminate the opening angle functions using the statistical tools available to us. This is illustrated in Table \ref{tab.3}.

\begin{table}
 \centering
  \caption{Number of good solutions for different combination of distribution functions described by Equations: (\ref{eq.cosGH}) -- inclination angle, (\ref{eq.lognor}) -- log--normal period distribution, (\ref{eq.gamma2}) -- gamma period distribution; (\ref{eq.rho.biggs}), (\ref{eq.rho.lm}) and (\ref{eq.rho.gks}b) -- opening angle distribution functions.
\label{tab.3}}
  \begin{tabular}{|c|l|c|}
\hline
\multicolumn{1}{|c|}{No.} & \multicolumn{1}{|l|}{Distributions} & \multicolumn{1}{|c|}{No. of solutions}\\\hline\hline
 1  & (\ref{eq.cosGH}) (\ref{eq.lognor}) (\ref{eq.rho.lm})    & 326\\
 2  & (\ref{eq.cosGH}) (\ref{eq.lognor}) (\ref{eq.rho.biggs}) & 170\\
 3  & (\ref{eq.cosGH}) (\ref{eq.lognor}) (\ref{eq.rho.gks}b) & 183\\ \hline
 4  & (\ref{eq.cosGH}) (\ref{eq.gamma2}) (\ref{eq.rho.lm})    & 147\\
 5  & (\ref{eq.cosGH}) (\ref{eq.gamma2}) (\ref{eq.rho.biggs}) & 1 \\ \hline
\end{tabular}
\end{table}

\item [5.] The average beaming fraction $f_b=N_{det}/N_{tot}\approx$0.1, where $N_{tot}=50000$ is the total number of simulated pulsar candidates and $N_{det}$ is the number of detected pulsars in our simulations. This value is close to the one obtained by TM98 and ZJM03, but lower than those obtained by GH96 and KGM04. One should realise that this factor corresponds to averaging over many pulsar parameters, the most important being pulsar period (for more details see the end of this Section). Moreover, since the pulsar luminosity is not taken into account (some distant pulsars will be too weak to be detected) this value of the beaming fraction should be considered as an upper limit. Once the luminosity problem is considered (see Appendix \ref{sec.appendix.luminosity} for some details) the value of beaming fraction largely decreases, perhaps even to the value as low as 0.02.
\end{enumerate}

As already mentioned this paper is a continuation of our previous attempt to resolve the pulse--width statistics based on the modeling of pulsar geometry (KGM04). Although we improved the numbers and the quality of the observational databases as well as methods of data processing, we are still lacking an analysis of the possible effect of the intrinsic luminosity of radio pulsars on our results. This problem is, however, very difficult and complicated and we will discuss it only superficially, postponing a full treatment to a future paper (see also Appendix \ref{sec.appendix.luminosity}). The proper approach would be to compare the synthetic radio luminosity with the minimum detectable flux achieved in a given pulsar survey, and thus it can be applied only to uniform data sets of pulsars detected in single survey. Our data do not have such a degree of uniformity. However, most surveys were less sensitive to long--period pulsars, as it stems from the nature of the applied Fourier--transform method. Since the interpulse emission (which appears to be the most restrictive constrain in our analysis) occurs mainly at shorter periods (see upper panel in Fig. \ref{Fig.3}), a possible under--representation of pulsars with longer periods should not significantly affect our general results.

We assumed that the intrinsic pulsar luminosity does not depend significantly on the inclination angle (geometry) but can depend on $P$ and $\dot{P}$ (pulsar evolution). We performed simple test to check possible influence of pulsar geometry and evolution. The result of this test is presented and discussed in Appendix \ref{sec.appendix.luminosity}. However, the main results of this paper were obtained using only geometrical detection conditions, that is every pulsar in the field of view of our hypothetical radiotelescope was detected. It means that each detectable pulsar would be close and bright enough so its radio energy flux would exceed the detection threshold of our hypothetical radio pulsar search campaign. As we argue below (see also Appendix \ref{sec.appendix.luminosity}), omitting the luminosity problem does not affect correctness of our results, at least significantly. However, taking it under consideration would vastly complicate the research programme (especially its computational part) and introduce additional uncertainties decreasing reliability of our conclusions. Also, it is worth emphasizing once more that our statistical studies were in principle limited to geometrical features (inclination angles, pulse--width and structure of the beam), without drawing conclusions depending on the luminosity problem, like the initial periods or the neutron stars birth rates.

The absolute luminosity of pulsar can be written in general form as $L_r=f(P, \dot{P}) = A P^{\alpha_1} \dot{P}^{\alpha_2}$, where A, $\alpha_1$ and $\alpha_2$ are free parameters of a model (e.g. Arzoumanian, Chernoff \& Cordes, 2002; ACC02 hereafter). Some authors even argue that for given $P$ and $\dot{P}$ the luminosity is constant i.e. pulsars can be considered as standard candles. Without a detailed discussion we would like to emphasise that each particular approach depends on an adopted model and leads to some problems. For example, from the analysis of observational data \citet{acc02} deduced that luminosity function is $L_r=P^{-1.3} \dot{P}^{0.4}/10^{-15} 10^{29.3}$ erg s$^{-1}$. This could be considered as an empirical function, but one should realise that the authors used some controversial  assumptions about the shape and size of the beam to obtain it, which are not really true in our opinion. However, more important are problems reported by other authors \citep{gonthier04}, who tried to use this function in their statistical studies. They noticed that this function gives too many bright pulsars as compared with observations and too high the NS birth rate. For example, \citet{gonthier04} used the above function in simulations of $\gamma$--ray pulsars detected by EGRET \citep{thompson08} but they had to reduce the amplitude $A=10^{29.3}$ by factor of 60 (leaving values of ${\alpha_1}$ and ${\alpha_2}$ unchanged for unknown reasons). Anyway, it is important to realize that a form of the function of absolute pulsar luminosity is not known.

The flux density of a pulsar with luminosity $L$ erg s$^{-1}$ and distance $d$ is $S=(f_b/4\pi)L$ [mJy kpc$^2$], where $f_b$ is the beaming fraction describing a part of a full sphere illuminated by a pulsar beam. This value of $S$, calculated with high uncertainty, should be compared with the sensitivity of a radiotelescope expressed as the so-called minimum detectable flux density$S_{min}$. Detection of pulsar is possible if $S>S_{min}$. We used data from pulsars discovered in more than a dozen surveys, with $S_{min}$ changing from one campaign to another. It is impossible to find a universal, theoretical value of $S_{min}$ corresponding to all pulsars in our sample, an thus it is impossible to find a consistent detection criterion based only on a distance to the pulsar and its luminosity. Apart from uncertainties introduced by this method, the parameter space would be extended enormously by many dimensions such as: distance, direction in the space to the pulsar, inhomogeneities in ISM, luminosity function and beaming fraction. Thus, computationally the problem greatly complicates and the expected benefits are not that high to justify these efforts. Therefore, we used the assumption that each pulsar detected in our ''geometrical search'' simulations corresponds to a flux density $S$ greater than a hypothetical $S_{min}$, the value of which we do not specify. Instead, we use geometrical criterion $\rho_0 > \beta$ where $\rho = 1.1 \rho_{10}$ and $\rho_{10}$ is calculated from empirical formulae (Eqs. (\ref{eq.rho.lm}) -- (\ref{eq.rho.gks})). 

In Appendix \ref{sec.appendix.luminosity} we present a simple test to check how the luminosity problem can affect a validity of our conclusions. We do not intend to find any new model of the intrinsic pulsar luminosity. This is beyond the scope of our paper. Instead, we use a very convenient existing luminosity probability density function presented by \citet{rl10}, which is best suited for our non-uniform (with respect to sensitivity of different pulsar search campaigns) database. The results of this check are presented in Table \ref{tab.2} (lines 1b -- 9b in comparison with lines 1a -- 9a) and shortly explained below. We found that most of our conclusions were not significantly changed but the number of pulsar detections dropped by a factor of several (see Appendix \ref{sec.appendix.luminosity}). As a result the actual (but still period averaged) beaming fraction $f_b$ dropped to a value about 2\% (compared with 10\% obtained without luminosity).

Therefore, we conclude that the actual beaming fraction $f_b$ (describing a part of the full sphere illuminated by an average pulsar beam) can be significantly  lower than 10 \%, perhaps even as low as about 2\% (see Table \ref{tab.2}). This is by a factor of few to several smaller than estimates of the beaming fraction value that can be found in the literature (0.17 -- \citet{gunn70}, 0.16 -- GH96, 0.14 -- KGM04, 0.12 -- ZJM03, 0.1 -- TM98, 0.084 -- WJ08a). This would suggest that the number of the neutron stars in the Galaxy is much larger than currently estimated. Recent discoveries of Rotating Radio Transients \citep{mclaughlin06} seem to support such a point of view. Indeed, it seems that there are 2 -- 3 times more RRATs than pulsars in the Galaxy \citep{keane10b}. RRATs seem to be just a normal radio pulsars in which we detect only strongest subpulses. More than 30 RRATs have been detected so far \citep{keane10}, thus an interesting question arises when one should expect first interpulse case in those objects. Our paper shows that there are about 3\% of interpulses in normal pulsar population. Thus, among 30 RRATs one should expect one IP case. However, the situation is more complicated than that since in many cases the interpulse is much weaker than the main pulse. From careful analysis of all our interpulse cases we can predict that the first interpulse RRAT will occur when the population of these objects will rise to about 100. \\

Finally, we would like to comment on the new interpulse case detected in PSR J2007+2722 using Einstein@Home global computing technique \citep{knispel10}. This is a 24 ms isolated pulsar with magnetic axis almost aligned to the spin axis. The mean profile covers full 360$^{\circ}$ pulse window, with two equal amplitude components separated by about 180$^{\circ}$ (see Fig. 1 in \citet{knispel10}). Thus, this case seems to be an ideal example of SP-IP object. However, the authors of the discovery paper argue that this is likely a disrupted recycled pulsar. For this reason we do not include it to our interpulse database. The global computing technique should result in more pulsar discoveries in the near future and we hope that some of them will contain the interpulse emission.\\

\section{Conclusions}
The main results and conclusions of this paper can be summarized as follows:
\begin{enumerate}
\item [1.] We compiled the largest ever database of 44 pulsars with interpulses, in which we identified 31 (about 2\%) double--pole cases and 13 (about 1\%) single--pole cases. 
\item [2.] We found a strong evidence that the magnetic axis aligns with time towards the rotation axis from the originally random orientation.
\item [3.] The parent period distribution density function is most likely the log--normal distribution, although the gamma distribution cannot be excluded.
\item [4.] The most suitable model distribution function for the inclination angle is the complicated trigonometric function which has two local maxima, one near 0$^{\circ}$ and the other near 90$^{\circ}$. The former and the latter implies the right rates of IP occurrence, single--pole (almost aligned rotator) and double--pole (almost orthogonal rotator), respectively. 
\item [5.] The average pulsar beam has an almost perfect circular cross-section.
\item [6.] The upper limit for period averaged beaming fraction describing a fraction of the full sphere covered by pulsar beam is about 10\%. 
\end{enumerate}

\section*{Acknowledgments}
This work is partially supported by the Grant 1 P03D 015 29 of the Polish State Committee for Scientific Research and by Polish Research Grants NN 203 2738 33 and NN 203 3919 34. We thank R.N. Manchester for invitation of K.M. to Australia, support and access to the ATNF pulsar data. We also thank M. Kramer for sharing with us unpublished results concerning new interpulse cases. We thank M. Kolodziejczyk for linguistic help. We also thank the anonymous referee for useful comments.

\onecolumn
\twocolumn
\begin{center}
    \LARGE{ON--LINE MATERIALS}  \label{sec.appendix.online}
\end{center}
\appendix
\section{Luminosity problem}\label{sec.appendix.luminosity}
To test how the luminosity problem can affect the validity of results and conclusions obtained in our paper we followed the results of \citet{fk06}, recently presented  by Ridley \& Lorimer (2010; hereafter RL10). As shown in this paper the radio luminosity can be calculated from
\begin{equation}
log L = log L_0 - log P + 0.5\: log (\dot{P}/10^{-15})+\delta_L,
\label{eq.log.luminosity}
\end{equation}
where $L_0$ is 0.18 mJy kpc$^2$ and $\delta_L$ is randomly chosen from a normal distribution with $\sigma_{\delta_L}=0.8$. Thus the luminosity $L$ can be calculated when the values of $P$ and $\dot{P}$ are known. We generated pulsar period $P$ as a random deviate with a parent probability density function $f(P)$ corresponding to Eqs. (\ref{eq.gamma2}) -- (\ref{eq.lognor}) in Section \ref{sec.prob.density.funct}. The value of period derivative can be calculated from either Eqs. (9) and (10) in \citet{rl10} or by random generation with a distribution in agreement with the observed $P-\dot{P}$ diagram (Fig. \ref{Fig.6}). We used both methods and made sure that they lead to similar simulation results. For example, we found that $L$ depends very weakly on the inclination angle $\alpha$ which enters into Eq. (9) in \citet{rl10}.

\begin{figure}
\vspace{80pt}
\begin{center}
\includegraphics[width=8cm,height=5cm,angle=0]{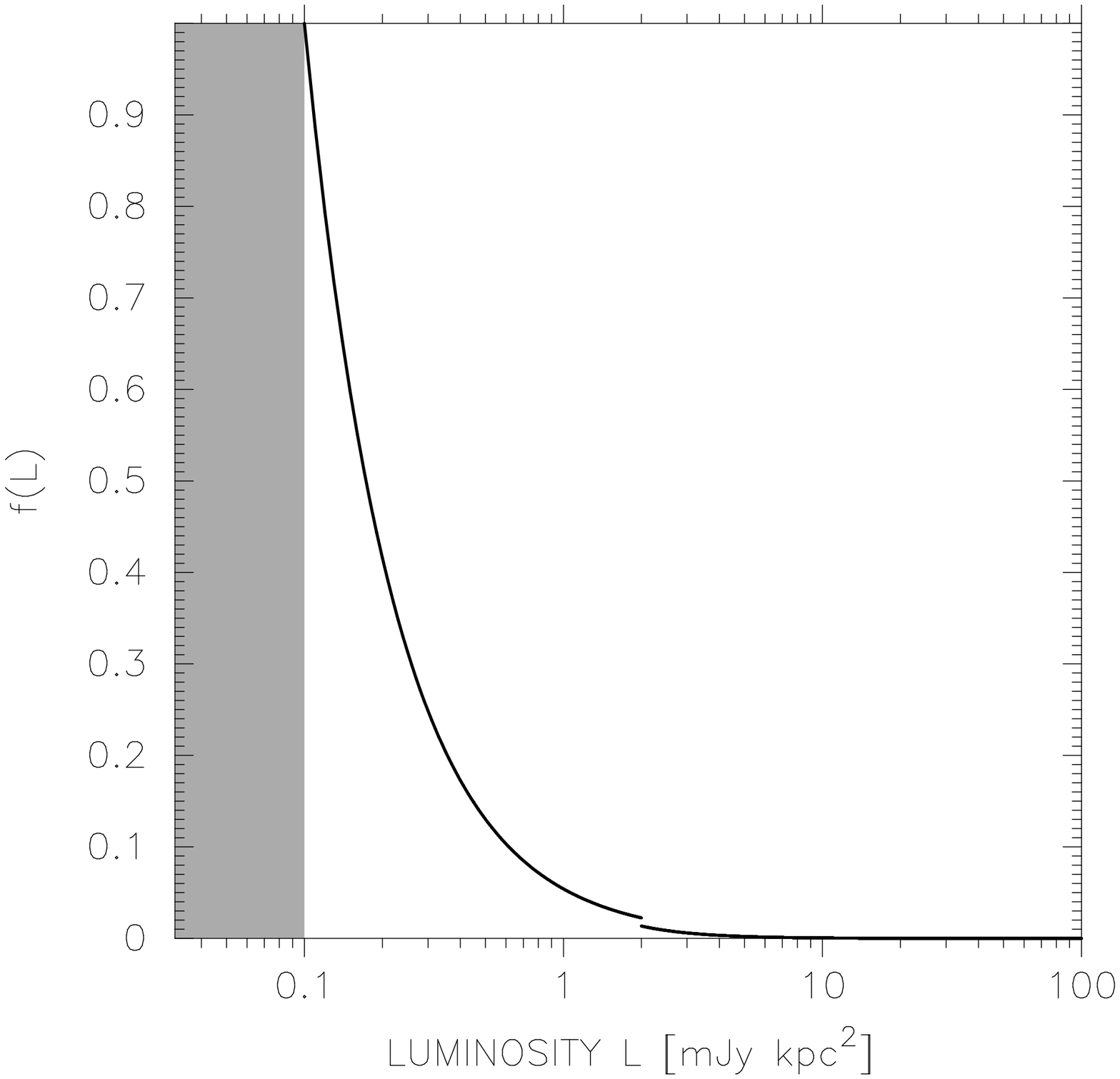}
\caption{The luminosity probability density function described by Eq. (\ref{eq.f.lum}). No pulsar detections are expected in shadowed rectangular area.
\label{Fig.A1}}
\end{center}
\end{figure}

Once the luminosity $L(P,\dot{P})$ is calculated we can obtain the probability density function
\begin{equation}
f(L) \propto \left\{
\begin{array}{l l l l l l}
0 & L \in [0 \; \text{mJy kpc}^2, 0.1\; \text{mJy kpc}^2)\\
L^{-19/15} & L \in [0.1 \; \text{mJy kpc}^2, 2.0\; \text{mJy kpc}^2)\\
L^{-2}     & L \in [2.0 \; \text{mJy kpc}^2, \infty\; \text{mJy kpc}^2),
\end{array} \right.
\label{eq.f.lum}
\end{equation}
(Eq.(17) in \citet{fk06} and Eq.(15) in \citet{rl10}). This function is illustrated in Fig. \ref{Fig.A1}. Now, we can extend our Monte Carlo simulation procedure by adding to the list of 16 steps listed in Section 4 the detection condition based on the luminosity described above. The procedure was as follows. First, to simplify calculations the function was normalised to unity at the luminosity value equal to 0.1 mJy kpc$^2$ (see Fig. \ref{Fig.A1}). This gave the proportionality constant $C=0.1^{19/15}=0.05$ in Eq. (\ref{eq.f.lum}). For a generated $P$ and $\dot{P}$ the values of $L$ and $f(L)$ were calculated according to Eqs. (\ref{eq.log.luminosity}) and (\ref{eq.f.lum}), respectively. The pulsar was counted as detected when the value of $f(L)$ lied below the thick line presented in Fig.\ref{Fig.A1} (and outside the shadowed area).

The results of the luminosity test are presented in Table \ref{tab.2}, where we compare good solutions with (lines 1b -- 9b) and without (lines 1a -- 9a) luminosity included, respectively. As one can see the only parameter that changed drastically is the beaming fraction $f_b$, which dropped by a factor of about 8. This is a consequence of much smaller number of detections. However, the interpulse detection at the requested levels still determines the statistical structure of detected pulsar population. This is illustrated in Fig. \ref{Fig.8}, as compared with Fig. \ref{Fig.7}.

In this Appendix we attempted to estimate an influence of the pulsar luminosity problem on the geometrical conclusion derived in our paper. We used a luminosity model represented by Eq. (\ref{eq.f.lum}) derived by \citet{fk06} and presented by \citet{rl10}. If this equation describes the luminosity density probability function for pulsars in our Galaxy, then the beaming fraction can be as low as about 2\%. The advantage of this rather crude model is a convenience in use and independence of details of different pulsar search campaigns. Anyway, at this point we can firmly conclude that the average pulsar beaming fraction is a number between 0.02 and 0.10 (2\% -- 10\%).

\onecolumn
\section{Tables}\label{sec.appendix.tables}

\begin{small}
\begin{longtable}{|c|c|c|c|c|c|c|}
\caption{Reproduction of the results from KGM04
\label{tab.kgm.results}}\\\hline

\multicolumn{4}{|l|}{Original results from KGM04}&  \multicolumn{3}{|l|}{Reproduction of KGM04 results in this work}\\ \hline

\multicolumn{1}{|l}{$m=0.35$} & \multicolumn{1}{l|}{$a=2.51$} & \multicolumn{5}{l|}{}\\   \hline

Parameter & Average & $\sigma$& $\%$ & Average & $\sigma$& $\%$ \\   \hline
       NOP            & 6385.3501&  70.9845 &        &   6399.0098 & 66.5424 & \\
      DP--IP           &  143.2600&  11.8651 &  2.244 &    141.2800 & 11.1410 &   2.208 \\
      SP--IP           &   41.5600&   6.7962 &  0.651 &     41.7100 &  6.4075 &   0.652 \\
$\mathcal{D}(W_{10})$ &    0.1270&   0.0066 &        &      0.1280 &  0.0060 &         \\
\hline
\end{longtable}
\end{small}

\vspace{1.5cm}

\begin{small}
\begin{longtable}{|r|l|r|r|r|r|r|r|r|r|r|r|}
\caption{Simulation results of the pulsar beam for opening angle distribution Eq. (\ref{eq.rho.lm}), inclination angle distribution function Eq. (\ref{eq.cosGH}) and period distribution function Eq. (\ref{eq.lognor}). Graphical representation of this class of solutions is represented in Fig. \ref{Fig.7}.
\label{tab.new.res.454f}}\\\hline
%\tiny
\multicolumn{1}{|r|}{No.} & \multicolumn{2}{|c|}{$f(P)$}& \multicolumn{2}{|c|}{$W_{10}$} &
\multicolumn{2}{|c|}{$P$} & \multicolumn{2}{|c|}{$\alpha$} & \multicolumn{1}{|c|}{\scriptsize{DP-IP}}& \multicolumn{1}{|c|}{\scriptsize{SP-IP}}&\multicolumn{1}{|c|}{$f_b$}\\ \cline{2-9}

\multicolumn{1}{|r|}{} & \multicolumn{1}{|c|}{$x_0$}& \multicolumn{1}{|c|}{$\sigma$} &
\multicolumn{1}{|c|}{$\mathcal{D}$} & \multicolumn{1}{|c|}{$\mathcal{P}$}& \multicolumn{1}{|c|}{$\mathcal{D}$} & \multicolumn{1}{|c|}{$\mathcal{P}$} & \multicolumn{1}{|c|}{$\mathcal{D}$} & \multicolumn{1}{|c|}{$\mathcal{P}$} &
\multicolumn{1}{|c|}{\footnotesize{[\%]}}& \multicolumn{1}{|c|}{\footnotesize{[\%]}}& \multicolumn{1}{|c|}{} \\ \hline

   1  &  -0.39  & 0.79  &  0.044  &  0.484  &  0.053  &  0.008  &  0.084  &  0.255  & 2.17  & 0.92  & 0.111  \\
   2  &  -0.38  & 0.74  &  0.041  &  0.533  &  0.051  &  0.007  &  0.089  &  0.191  & 2.21  & 1.00  & 0.111  \\
   3  &  -0.38  & 0.75  &  0.040  &  0.574  &  0.052  &  0.005  &  0.090  &  0.188  & 2.15  & 0.90  & 0.109  \\
   4  &  -0.38  & 0.76  &  0.039  &  0.593  &  0.049  &  0.016  &  0.087  &  0.218  & 2.00  & 0.89  & 0.110  \\
   5  &  -0.38  & 0.77  &  0.042  &  0.512  &  0.047  &  0.014  &  0.086  &  0.238  & 2.22  & 0.97  & 0.110  \\
\hline
\end{longtable}
\end{small}

\begin{small}
\begin{longtable}{|r|l|r|r|r|r|r|r|r|r|r|r|}
\caption{Simulation results of the pulsar beam for opening angle distribution Eq. (\ref{eq.rho.biggs}), inclination angle distribution function Eq. (\ref{eq.cosGH}) and period distribution function Eq. (\ref{eq.lognor}). Graphical representation of this class of solutions is represented in Fig. \ref{Fig.C3}.
\label{tab.new.res.354f}}\\\hline
\multicolumn{1}{|r|}{No.} & \multicolumn{2}{|c|}{$f(P)$}& \multicolumn{2}{|c|}{$W_{10}$} &
\multicolumn{2}{|c|}{$P$} & \multicolumn{2}{|c|}{$\alpha$} & \multicolumn{1}{|c|}{\scriptsize{DP-IP}}& \multicolumn{1}{|c|}{\scriptsize{SP-IP}}&\multicolumn{1}{|c|}{$f_b$}\\ \cline{2-9}

\multicolumn{1}{|r|}{} & \multicolumn{1}{|c|}{$x_0$}& \multicolumn{1}{|c|}{$\sigma$} &
\multicolumn{1}{|c|}{$\mathcal{D}$} & \multicolumn{1}{|c|}{$\mathcal{P}$}& \multicolumn{1}{|c|}{$\mathcal{D}$} & \multicolumn{1}{|c|}{$\mathcal{P}$} & \multicolumn{1}{|c|}{$\mathcal{D}$} & \multicolumn{1}{|c|}{$\mathcal{P}$} &
\multicolumn{1}{|c|}{\footnotesize{[\%]}}& \multicolumn{1}{|c|}{\footnotesize{[\%]}}& \multicolumn{1}{|c|}{} \\ \hline
   1  &  -0.29  & 0.75  &  0.087  &  0.008  &  0.051  &  0.005  &  0.086  &  0.235  & 2.22  & 1.04  & 0.112  \\
   2  &  -0.28  & 0.72  &  0.078  &  0.020  &  0.049  &  0.010  &  0.081  &  0.287  & 2.29  & 1.06  & 0.111  \\
   3  &  -0.28  & 0.73  &  0.076  &  0.027  &  0.049  &  0.009  &  0.079  &  0.313  & 2.38  & 1.03  & 0.111  \\
   4  &  -0.28  & 0.74  &  0.076  &  0.027  &  0.049  &  0.011  &  0.078  &  0.338  & 2.25  & 0.99  & 0.111  \\
   5  &  -0.28  & 0.78  &  0.089  &  0.005  &  0.048  &  0.014  &  0.084  &  0.253  & 2.21  & 1.06  & 0.113  \\
\hline
\end{longtable}
\end{small}

\begin{small}
\begin{longtable}{|r|l|r|r|r|r|r|r|r|r|r|r|}
\caption{Simulation results of the pulsar beam for opening angle distribution Eq. (\ref{eq.rho.gks}b), inclination angle distribution function Eq. (\ref{eq.cosGH}) and period distribution function Eq. (\ref{eq.lognor}). Graphical representation of this class of solutions is represented in Fig. \ref{Fig.C4}.
\label{tab.new.res.554f}}\\\hline
%\tiny
\multicolumn{1}{|r|}{No.} & \multicolumn{2}{|c|}{$f(P)$}& \multicolumn{2}{|c|}{$W_{10}$} &
\multicolumn{2}{|c|}{$P$} & \multicolumn{2}{|c|}{$\alpha$} & \multicolumn{1}{|c|}{\scriptsize{DP-IP}}& \multicolumn{1}{|c|}{\scriptsize{SP-IP}}&\multicolumn{1}{|c|}{$f_b$}\\ \cline{2-9}

\multicolumn{1}{|r|}{} & \multicolumn{1}{|c|}{$x_0$}& \multicolumn{1}{|c|}{$\sigma$} &
\multicolumn{1}{|c|}{$\mathcal{D}$} & \multicolumn{1}{|c|}{$\mathcal{P}$}& \multicolumn{1}{|c|}{$\mathcal{D}$} & \multicolumn{1}{|c|}{$\mathcal{P}$} & \multicolumn{1}{|c|}{$\mathcal{D}$} & \multicolumn{1}{|c|}{$\mathcal{P}$} &
\multicolumn{1}{|c|}{\footnotesize{[\%]}}& \multicolumn{1}{|c|}{\footnotesize{[\%]}}& \multicolumn{1}{|c|}{} \\ \hline
   1  &  -0.38  & 0.78  &  0.063  &  0.104  &  0.052  &  0.005  &  0.080  &  0.302  & 2.27  & 1.02  & 0.113  \\
   2  &  -0.37  & 0.75  &  0.057  &  0.194  &  0.051  &  0.006  &  0.083  &  0.267  & 2.23  & 1.03  & 0.112  \\
   3  &  -0.37  & 0.76  &  0.062  &  0.109  &  0.051  &  0.006  &  0.085  &  0.242  & 2.25  & 1.06  & 0.112  \\
   4  &  -0.37  & 0.77  &  0.062  &  0.114  &  0.051  &  0.009  &  0.082  &  0.285  & 2.22  & 1.08  & 0.112  \\
   5  &  -0.37  & 0.78  &  0.065  &  0.088  &  0.051  &  0.005  &  0.083  &  0.266  & 2.19  & 1.01  & 0.113  \\
\hline
\end{longtable}
\end{small}

\begin{small}
\begin{longtable}{|r|l|r|r|r|r|r|r|r|r|r|r|}
\caption{Simulation results of the pulsar beam for opening angle distribution Eq. (\ref{eq.rho.lm}), inclination angle distribution function Eq. (\ref{eq.cosGH}) and period distribution function Eq. (\ref{eq.gamma2}). Graphical representation of this class of solutions is represented in Fig. \ref{Fig.C6}.
\label{tab.new.res.451f}}\\\hline
\multicolumn{1}{|r|}{No.} & \multicolumn{2}{|c|}{$f(P)$}& \multicolumn{2}{|c|}{$W_{10}$} &
\multicolumn{2}{|c|}{$P$} & \multicolumn{2}{|c|}{$\alpha$} & \multicolumn{1}{|c|}{\scriptsize{DP-IP}}& \multicolumn{1}{|c|}{\scriptsize{SP-IP}}&\multicolumn{1}{|c|}{$f_b$}\\ \cline{2-9}

\multicolumn{1}{|r|}{} & \multicolumn{1}{|c|}{$x_0$}& \multicolumn{1}{|c|}{$\sigma$} &
\multicolumn{1}{|c|}{$\mathcal{D}$} & \multicolumn{1}{|c|}{$\mathcal{P}$}& \multicolumn{1}{|c|}{$\mathcal{D}$} & \multicolumn{1}{|c|}{$\mathcal{P}$} & \multicolumn{1}{|c|}{$\mathcal{D}$} & \multicolumn{1}{|c|}{$\mathcal{P}$} &
\multicolumn{1}{|c|}{\footnotesize{[\%]}}& \multicolumn{1}{|c|}{\footnotesize{[\%]}}& \multicolumn{1}{|c|}{} \\ \hline
   1  &   0.37  & 2.32  &  0.036  &  0.687  &  0.049  &  0.008  &  0.088  &  0.206  & 2.13  & 0.96  & 0.110  \\
   2  &   0.37  & 2.33  &  0.038  &  0.634  &  0.050  &  0.005  &  0.090  &  0.189  & 2.18  & 0.94  & 0.110  \\
   3  &   0.37  & 2.35  &  0.035  &  0.729  &  0.049  &  0.007  &  0.091  &  0.177  & 2.07  & 0.92  & 0.109  \\
   4  &   0.37  & 2.36  &  0.036  &  0.690  &  0.050  &  0.006  &  0.090  &  0.181  & 2.06  & 0.89  & 0.109  \\
   5  &   0.37  & 2.37  &  0.032  &  0.797  &  0.049  &  0.009  &  0.086  &  0.231  & 2.21  & 0.98  & 0.109  \\
\hline
\end{longtable}
\end{small}

\begin{small}
\begin{longtable}{|r|l|r|r|r|r|r|r|r|r|r|r|}
\caption{Simulation results of the pulsar beam for opening angle distribution Eq. (\ref{eq.rho.biggs}), inclination angle distribution function Eq. (\ref{eq.cosGH}) and period distribution function Eq. (\ref{eq.gamma2}). Graphical representation of this class of solutions is represented in Fig. \ref{Fig.C5}.
\label{tab.new.res.351f}}\\\hline
\multicolumn{1}{|r|}{No.} & \multicolumn{2}{|c|}{$f(P)$}& \multicolumn{2}{|c|}{$W_{10}$} &
\multicolumn{2}{|c|}{$P$} & \multicolumn{2}{|c|}{$\alpha$} & \multicolumn{1}{|c|}{\scriptsize{DP-IP}}& \multicolumn{1}{|c|}{\scriptsize{SP-IP}}&\multicolumn{1}{|c|}{$f_b$}\\ \cline{2-9}

\multicolumn{1}{|r|}{} & \multicolumn{1}{|c|}{$x_0$}& \multicolumn{1}{|c|}{$\sigma$} &
\multicolumn{1}{|c|}{$\mathcal{D}$} & \multicolumn{1}{|c|}{$\mathcal{P}$}& \multicolumn{1}{|c|}{$\mathcal{D}$} & \multicolumn{1}{|c|}{$\mathcal{P}$} & \multicolumn{1}{|c|}{$\mathcal{D}$} & \multicolumn{1}{|c|}{$\mathcal{P}$} &
\multicolumn{1}{|c|}{\footnotesize{[\%]}}& \multicolumn{1}{|c|}{\footnotesize{[\%]}}& \multicolumn{1}{|c|}{} \\ \hline
   1  &   0.38  & 2.51  &  0.071  &  0.047  &  0.049  &  0.009  &  0.083  &  0.266  & 2.29  & 1.07  & 0.110  \\
\hline
\end{longtable}
\end{small}

\twocolumn

\section{Figures}\label{sec.appendix.figures}

\begin{figure}
\vspace{40pt}
\begin{center}
\includegraphics[width=7.5cm,height=3.5cm,angle=0]{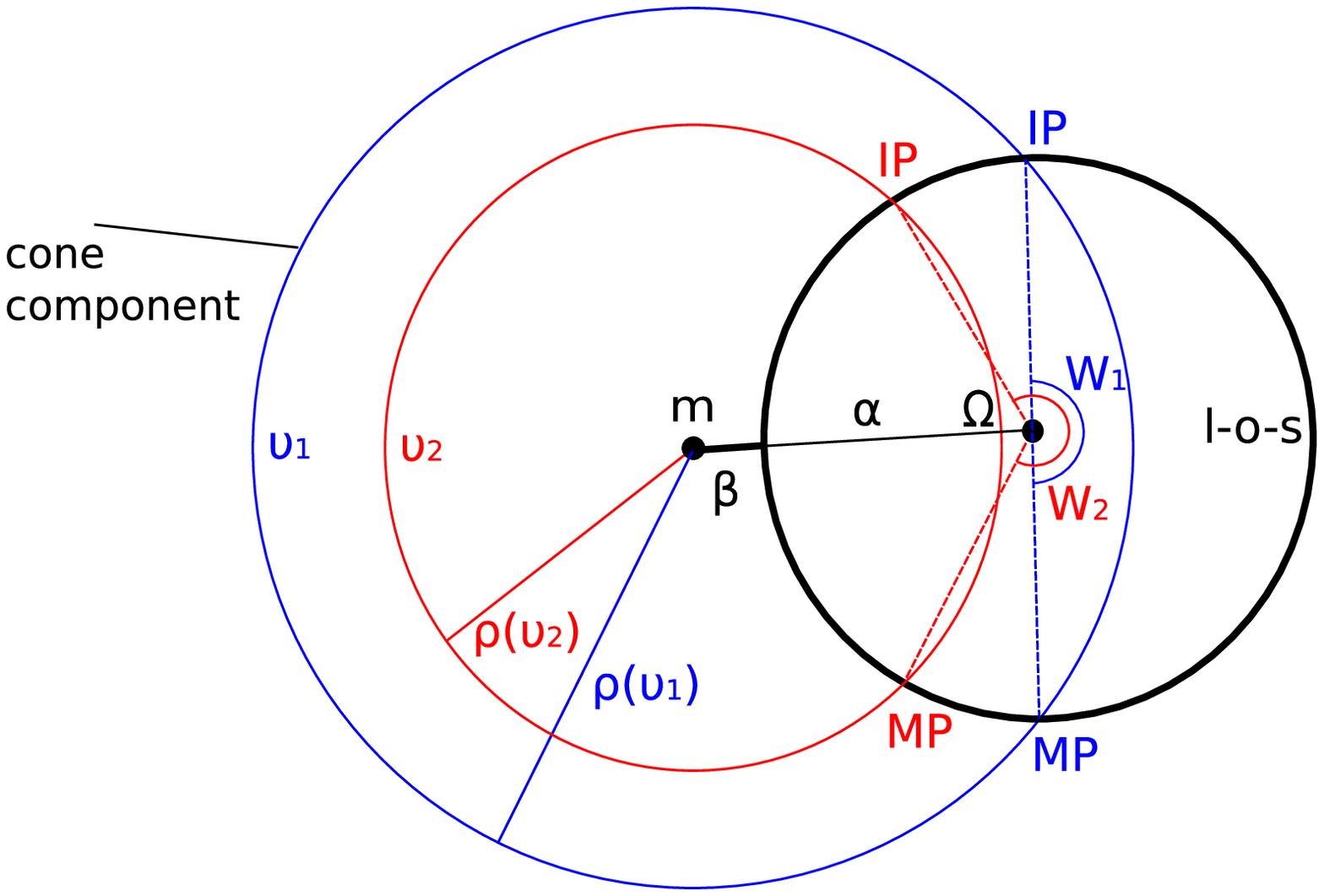}
\caption{Model of the single--pole interpulse -- ML77 version. Two cones of average emission corresponding to frequencies $\nu_1$ (blue) and $\nu_2 > \nu_1$ (red) with an opening angle $\rho(\nu_1) > \rho(\nu_2)$ are marked schematically. For a given $\alpha$ and $\beta$ angles the l--o--s (marked by the black circle) cuts the radiation cones twice. In this figure it is assumed that at frequency $\nu_1$ MP and IP are separated about $W_1 = 180^{\circ}$ of longitude. However with increasing frequency the separation $W$ will be different ($W_2 > W_1$ for the presented geometry). This frequency dependence of MP--IP separation is the main difference between ML77 and G83 (presented in Fig. \ref{Fig.C2}) versions of SP--IP models.
\label{Fig.C1}}
\end{center}
\end{figure}

\begin{figure}
\vspace{30pt}
\begin{center}
\includegraphics[width=6cm,height=4cm,angle=0]{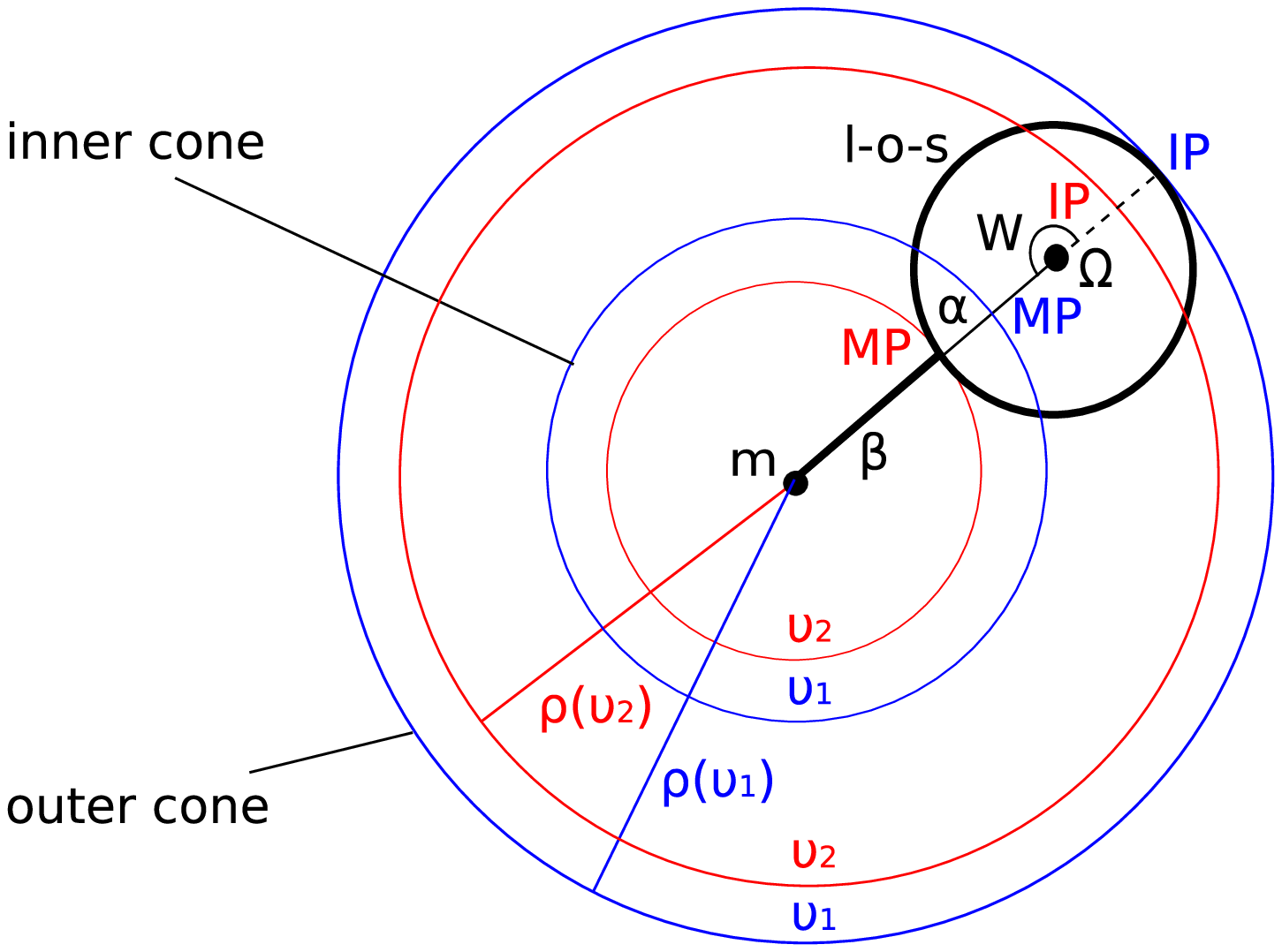}
\caption{Model of the single--pole interpulse for two nested hollow cones (or inner core surrounded by the cone) -- G83 version. In this case the mean pulsar beam consist of core and cone or two nested, coaxial cones (blue circles for frequency $\nu_1$ and red circles of frequency $\nu_2 > \nu_1$). For a given $\alpha$ and $\beta$ angles the l--o--s (black circle) cuts through both cones. The inner and the outer cone is responsible for the occurrence of the MP and the IP, respectively. For different (higher) observational frequency $\nu_2$ (red circles) the opening angle $\rho(\nu_1) > \rho(\nu_2)$. However, this version of the SP--IP model the MP--IP separation $W$ is frequency independent and equal to $180^{\circ}$ of longitude.
\label{Fig.C2}}
\end{center}
\end{figure}

\begin{figure}
\vspace{5pt}
\begin{center}
\includegraphics[width=8cm,height=8cm,angle=0]{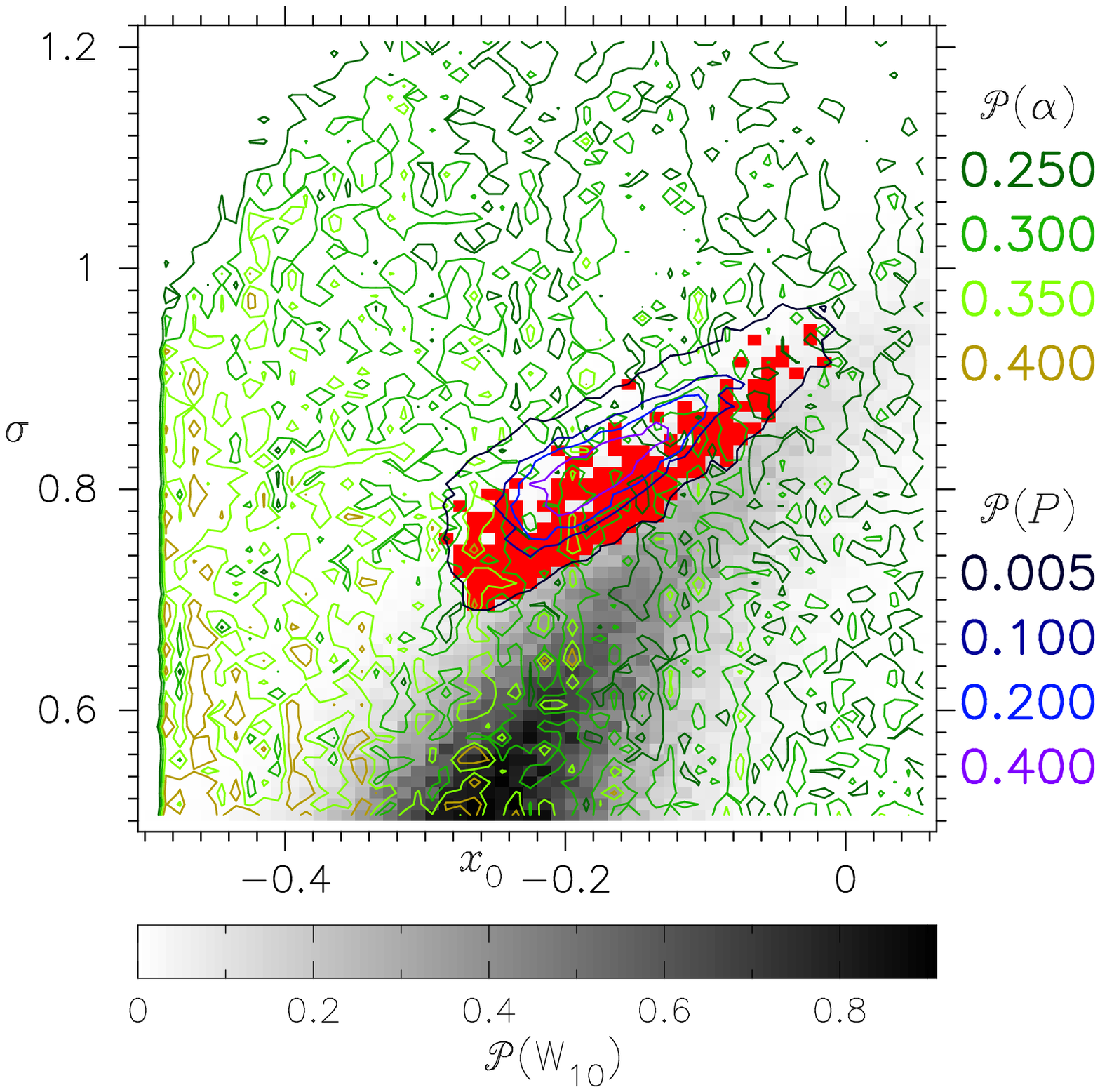}%354f
\caption{As in Fig. \ref{Fig.7} but for opening angles described by Eq. (\ref{eq.rho.biggs}). All acceptable solutions are represented by red rectangles and their distribution describes errors of the log--normal functions $x_0=-0.18^{+0.16}_{-0.11}$ and $\sigma=0.81^{+0.13}_{-0.12}$ (see also Tab.\ref{tab.new.res.354f}).
\label{Fig.C3}}
\end{center}
\end{figure}

\begin{figure}
\vspace{5pt}
\begin{center}
\includegraphics[width=8cm,height=8cm,angle=0]{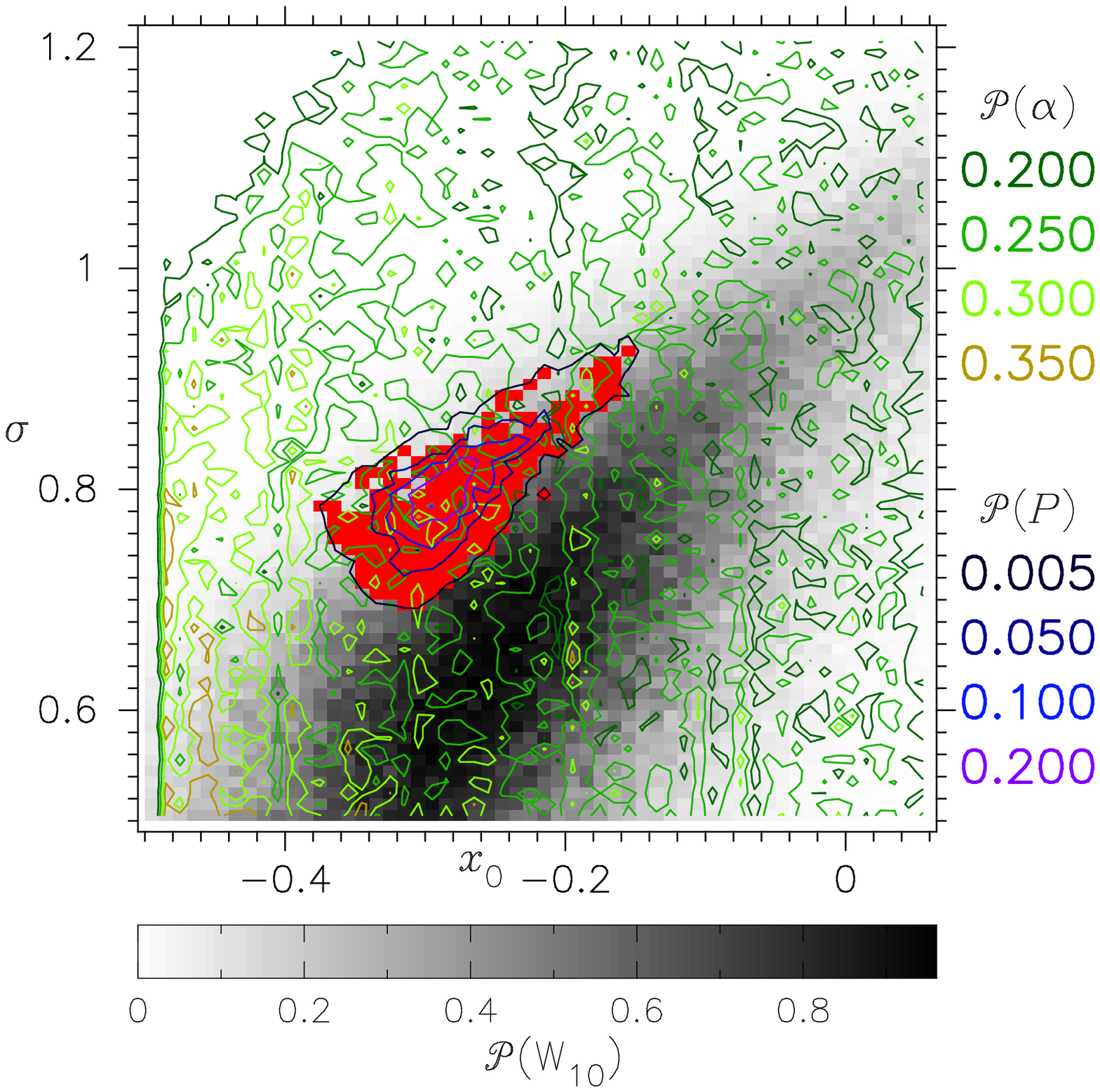}%554f
\caption{As in Fig. \ref{Fig.7} but for opening angles described by Eq. (\ref{eq.rho.gks}b). All acceptable solutions are represented by red rectangles and their distribution describes errors of the log--normal functions $x_0=-0.29^{+0.13}_{-0.09}$ and $\sigma=0.79^{+0.13}_{-0.10}$ (see also Tab.\ref{tab.new.res.554f}).
\label{Fig.C4}}
\end{center}
\end{figure}

\clearpage

\begin{figure}
\vspace{5pt}
\begin{center}
\includegraphics[width=8cm,height=8cm,angle=0]{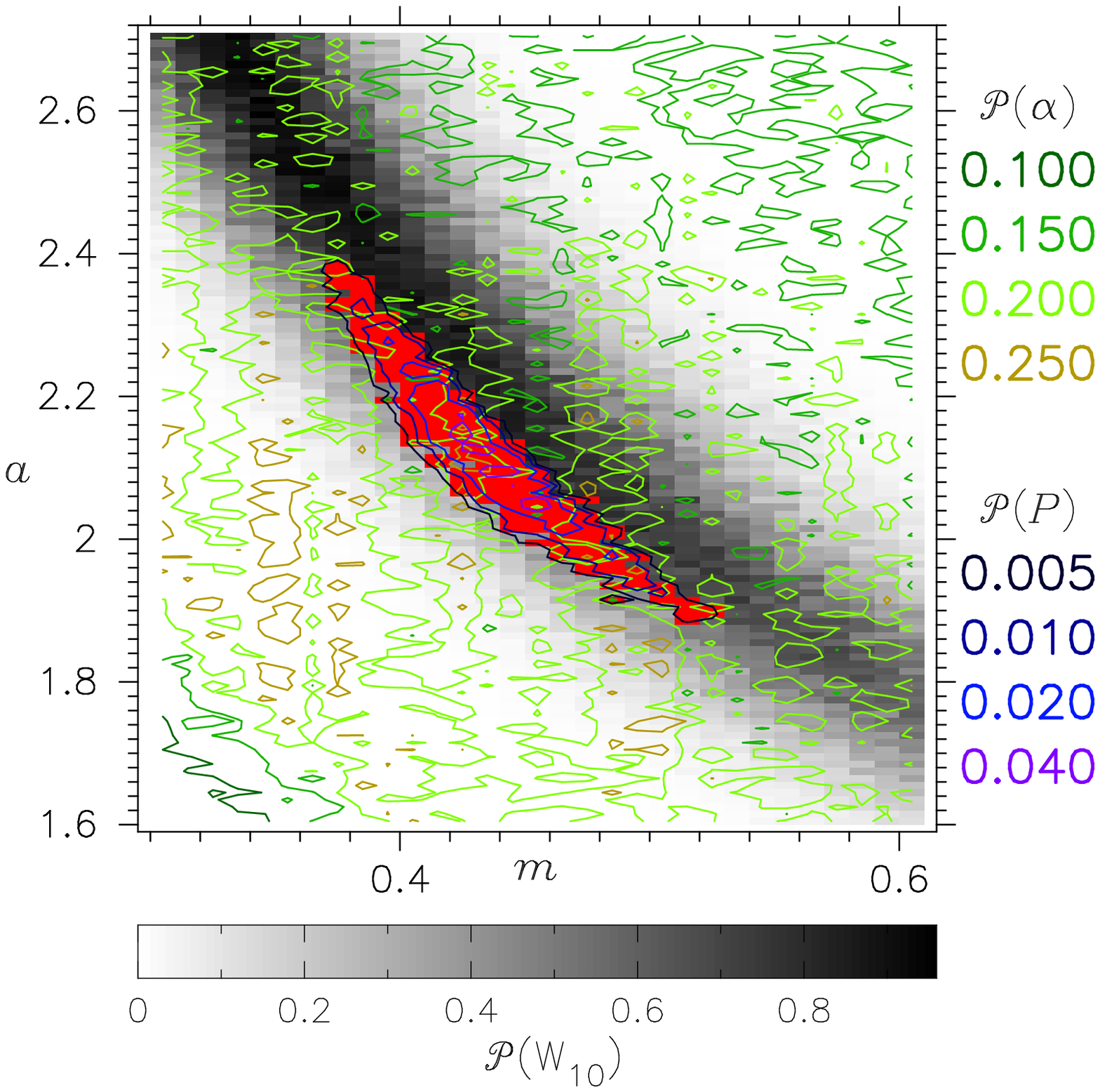}%451f
\caption{As in Fig. \ref{Fig.7} but for period distribution function Eq. (\ref{eq.gamma2}) and opening angles described by Eq. (\ref{eq.rho.lm}). All acceptable solutions are represented by red rectangles and their distribution describes errors of the gamma functions $m=0.45^{+0.07}_{-0.08}$ and $a=2.04^{+0.34}_{-0.16}$ (see also Tab.\ref{tab.new.res.451f}).
\label{Fig.C5}}
\end{center}
\end{figure}

\begin{figure}
\vspace{5pt}
\begin{center}
\includegraphics[width=8cm,height=8cm,angle=0]{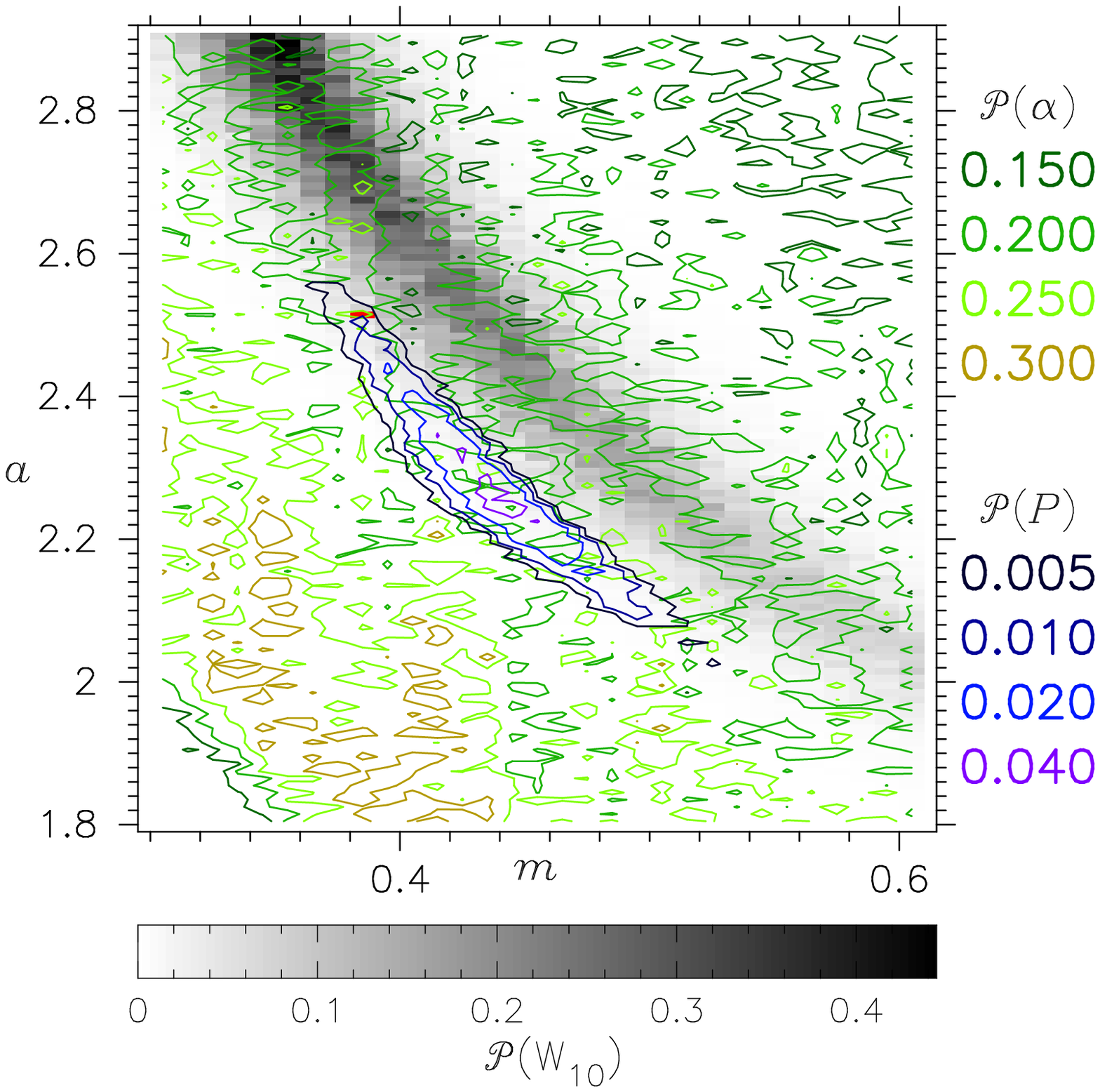}%351f
\caption{As in Fig. \ref{Fig.C5} but for opening angles described by Eq. (\ref{eq.rho.biggs}). The only one acceptable solution is represented by red rectangle with the gamma function parameters $m=0.38$ and $a=2.51$ (see also Tab.\ref{tab.new.res.351f}).
\label{Fig.C6}}
\end{center}
\end{figure}

\begin{figure}
\vspace{5pt}
\begin{center}
\includegraphics[width=8cm,height=8cm,angle=0]{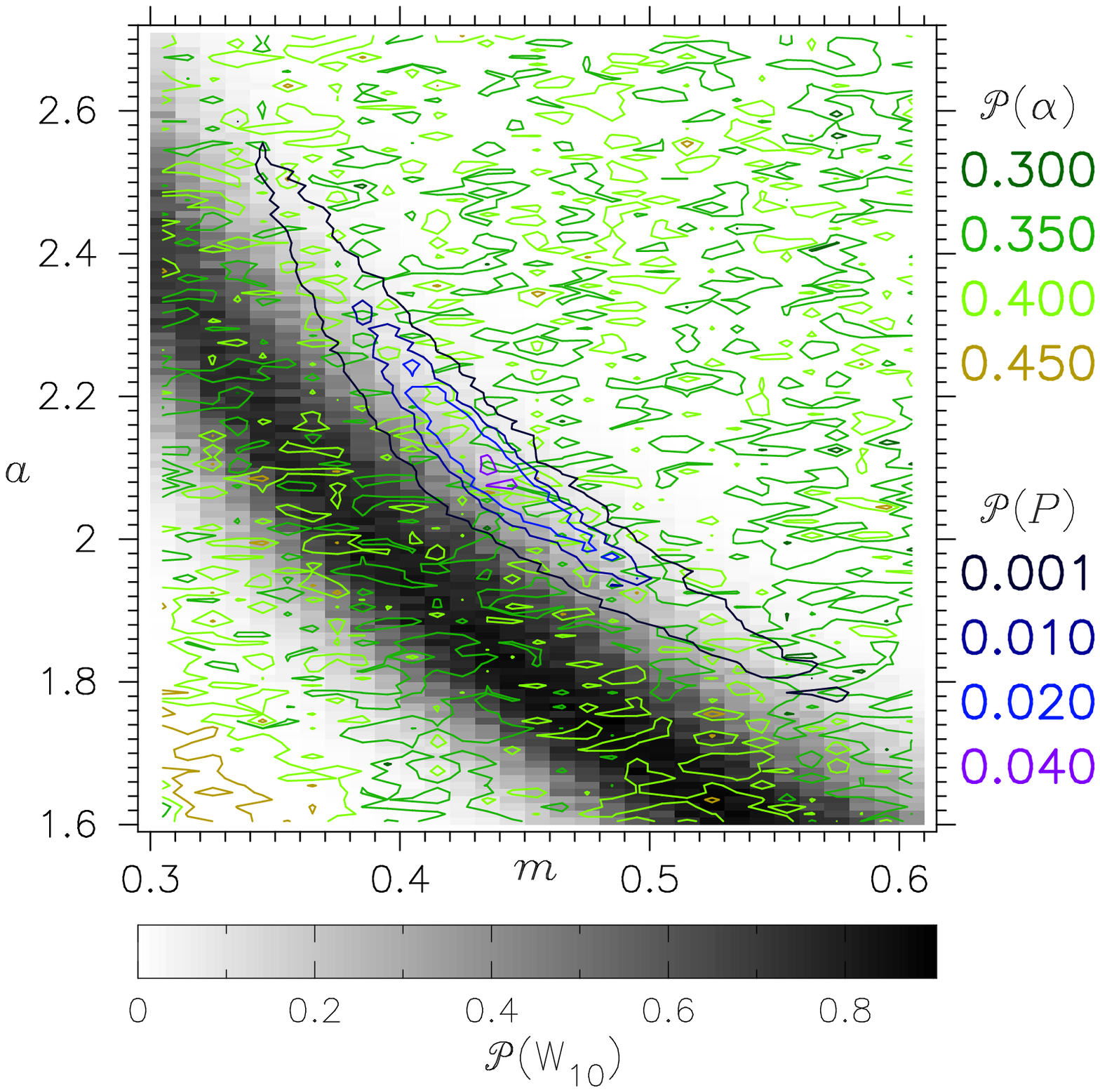}%441f
\caption{As in Fig. \ref{Fig.7} but for period distribution function Eq. (\ref{eq.cosZJM}) and opening angles described by Eq. (\ref{eq.rho.lm}). Notice that no interpulses were found.
\label{Fig.C7}}
\end{center}
\end{figure}

\begin{figure}
\vspace{5pt}
\begin{center}
\includegraphics[width=8cm,height=8cm,angle=0]{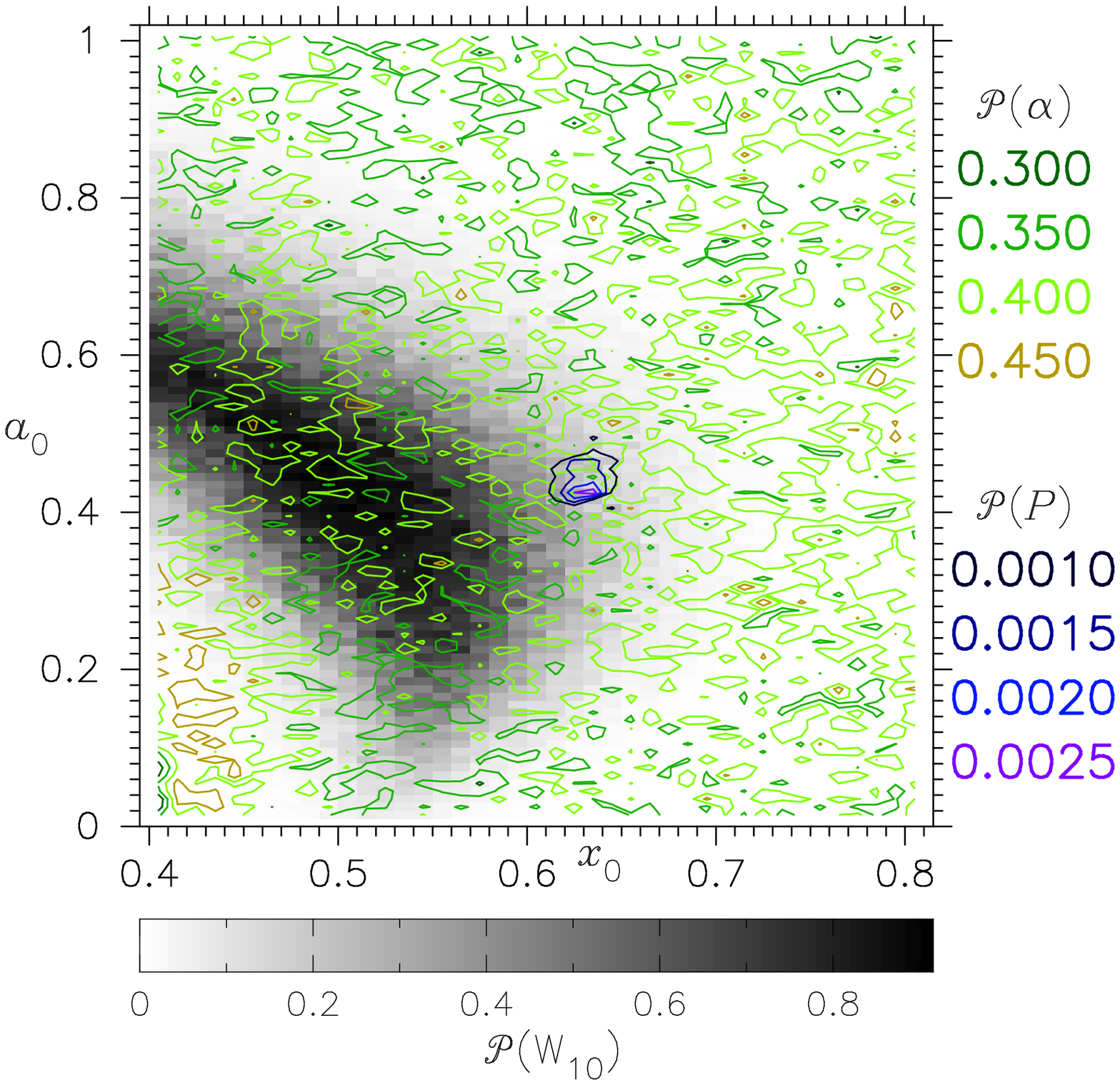}%442f
\caption{As in Fig. \ref{Fig.C7} but for period distribution function Eq. (\ref{eq.lorentz}).Notice that no interpulses were found.
\label{Fig.C8}}
\end{center}
\end{figure}.

\clearpage

\begin{figure}
\vspace{5pt}
\begin{center}
\includegraphics[width=8cm,height=8cm,angle=0]{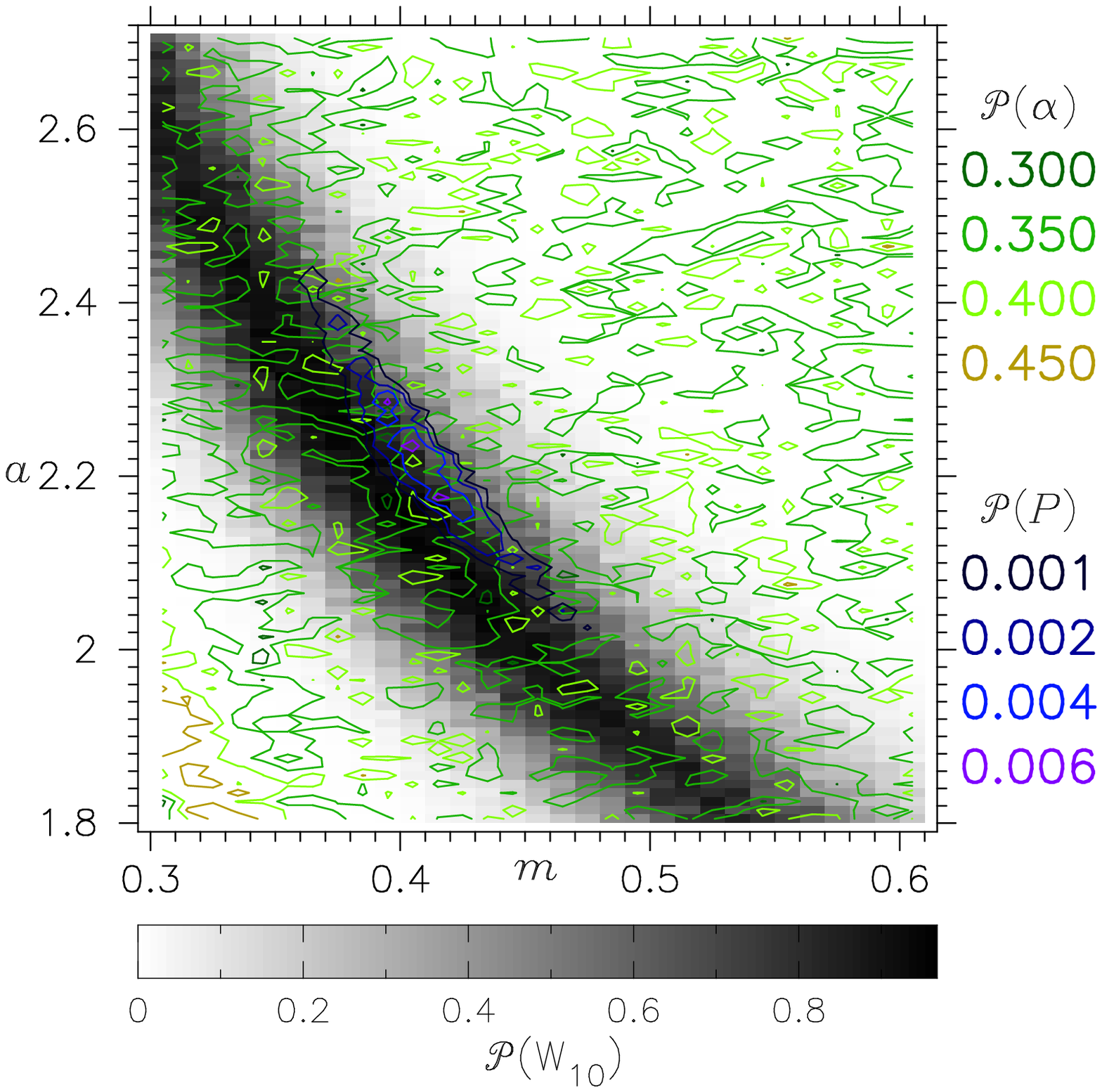}%541f
\caption{As in Fig. \ref{Fig.C7} but for opening angles described by Eq. (\ref{eq.rho.gks}b). Notice that no interpulses were found.
\label{Fig.C9}}
\end{center}
\end{figure}

\bsp
\label{lastpage}

\end{document}